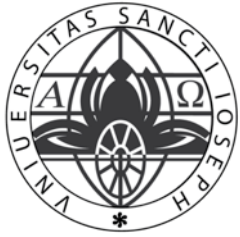

# UNIVERSITY OF SAINT JOSEPH
## 聖若瑟大學

GROKYA:

A PRIVACY-FRIENDLY FRAMEWORK FOR UBIQUITOUS

COMPUTING

A Thesis

Presented to

The Academic Faculty

by

**Daniel Filipe G. Farinha**

In Partial Fulfillment of the Requirements for the Degree of

Master in Information Technology

in the Faculty of Creative Industries

University of Saint Joseph, Macau

October 2015

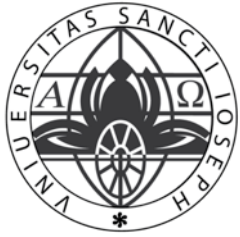

# UNIVERSITY OF SAINT JOSEPH
## 聖若瑟大學

**ENDORSEMENT**

| I certify that this report is solely my work, and that it has never been previously submitted for any academic award. | I, the supervisor, believe that this Dissertation is ready for assessment, and reaches the accepted standard for the Master in Information Technology. |
|---|---|
| Daniel Filipe G. Farinha | Dr. Neena Thota |

# ABSTRACT

In a world where for-profit enterprises are increasingly looking to maximize profits by engaging in privacy invading consumer-profiling techniques, the rise of ubiquitous computing and the Internet of Things (IoT) poses a major problem. If not acted upon quickly, the combination of Big Data with IoT will explode into a dystopian world that even George Orwell could not have predicted.

The proposed project aims to fill a gap that no other solution is addressing, which is to reach a win-win scenario that works for both the enterprises and the consumers. It aims to do this by creating the building blocks for a consumer owned infrastructure that can provide both privacy for the user, and still enable the enterprises to achieve their high-level goals.

# ACKNOWLEDGEMENTS

Firstly I would like to thank my thesis supervisor Prof. Neena Thota, for tirelessly attempting to keep me focused and on track, and for always providing prompt help whenever solicited.

My gratefulness goes to the academic staff of the Faculty of Creative Industries, especially Prof. Álvaro Barbosa, Neena Thota, João Garrot, and Gerald Estadieu who organised and taught the academic modules for this program. I am very lucky to have you as both teachers and colleagues. A thank you also goes to Profs Luís Gustavo Martins and Hsing Mei who, as visiting professors, delivered great classes. Still within the faculty, a special thank you to João Cordeiro and Filipa Martins for inspiring me with their work, creativity, and vision.

I must acknowledge my colleagues at the Office of Technology, Dora Pires, Guillaume Leclerc, Kitty Chan, Marco Santos, Teddy Cheong and Marco Leong, for contributing to a great work environment. I am truly fortunate to have them as colleagues. A special thank you must go to Marco Santos for designing the logo of one of Grokya's offshoot projects, subsatoshi.org.

My special gratitude goes to Gerald Estadieu for helping me with 3D printing and Marco Leong for helping with sourcing some of the components used in this project. Also thanks to both of them for patiently listening to my regular brainstorming about Grokya over the course of the past 6 years.

I am grateful to my parents back in Portugal for always supporting me, and to my new family in Macau for welcoming me and making me feel at home.

Finally I must thank my wife Geraldina and my baby daughter Lara for forgiving my absence over the past months as the thesis work intensified. Daddy is back!

**TABLE OF CONTENTS**

# LIST OF TABLES

# LIST OF FIGURES

# LIST OF ABBREVIATIONS

| | | |
|---|---|---|
| API | - | Application Programming Interface |
| BLE | - | Bluetooth Low Energy |
| BMI | - | Brain-Machine Interfaces |
| CRSP | - | Cascading Revenue Sharing Protocol |
| DNS | - | Domain Name System |
| DDNS | - | Dynamic Domain Name System |
| FEC | - | Forward Error Correction |
| HDD | - | Hard Disk Drive |
| HTTP | - | Hypertext Transfer Protocol |
| HTTPS | - | Hypertext Transfer Protocol Secure |
| IoT | - | Internet of Things |
| GPL | - | General Public License |
| GSM | - | Global System for Mobile (Communications) |
| LS | - | LifeServer |
| KYC | - | Know Your Customer |
| LBS | - | Location-Based System/Service |
| M2M | - | Machine-to-Machine |
| OHMD | - | Optical Head-Mounted Display |
| PC | - | Personal Computer |
| PDS | - | Personal Data Store |
| PoC | - | Proof of Concept |
| PoV | - | Point of View |
| PRNG | - | Pseudo-Random Number Generator |
| NFC | - | Near-Field Communications |
| SDK | - | Software Development Kit |
| SSL | - | Secure Sockets Layer |
| TLS | - | Transport Layer Security |
| TRNG | - | True Random Number Generator |
| VDP | - | Value Distribution Protocol |

# 1. INTRODUCTION

This project as a concept dates back to 2007, when the author was contemplating the future of mobile marketing and how behavioural advertising and profiling by a smartphone were clearly a winning strategy. From advertising the author quickly jumped to context-aware computing, in which ads are just another type of content that can be adapted to the user's context.

At the time the author wrongly assumed that consumers cared enough about their privacy as to not allow any kind of profiling my mobile handsets, so a privacy protecting solution was needed. Works out that consumers do not care as much as we think, however more recent developments in the areas of Big Data and the Internet of Things (IoT) are more than enough justification to work on a privacy-protecting alternative.

This project involves the engineering of consumer-side infrastructure, in the shape of a small personal server that will allow the users to engage in self-profiling, opening the way for higher-level interactions with enterprises without sharing private data.

The project was developed in a prototyping fashion, with an initial focus on the hardware. Future work will build upon this, by creating a peer-to-peer network of consumer devices that can work together to achieve the same high-level goals of Big Data analytics without incurring privacy issues for the consumers.

This thesis is just the beginning of a long journey. Fasten your seat belts.

## 2. LITERATURE REVIEW

### 2.1. INTRODUCTION

Ubiquitous computing in general, and context-aware computing in particular, have the potential to revolutionize the way humans interact with their environment, by allowing technology to adapt to individuals and their constantly changing needs. As will be shown section 2.2, ubiquitous computing can even make possible unique new benefits to society that otherwise would be difficult or impossible to achieve.

Unfortunately that same ecosystem of technologies generates a vast amount of personal data and carries a significant risk in the form of invasion of privacy. This is due to both a lack of security in the technology stack as well as economic motives that lead to explicit privacy intrusion by third parties, in what is now commonly referred to as Big Data Analytics. This problem cannot be understated. To support this view section 2.3.2 includes enough research evidence that should scare away even the strongest of ubiquitous computing enthusiasts.

If invasion of privacy is recognized as a major problem, not only from ubiquitous computing but also from other recent technological, economic and political developments, then it is only natural that efforts have been made to address the issue. Some of these efforts are covered in section 2.4.

Unfortunately despite well-intentioned efforts, strong economic and political forces continue to drive society towards an Orwellian future without privacy.

## 2.2. UBIQUITOUS AND CONTEXTUAL COMPUTING

> *"The most profound technologies are those that disappear. They weave themselves into the fabric of everyday life until they are indistinguishable from it."* (Weiser, 1991)

### 2.2.1. UBIQUITOUS COMPUTING

Ubiquitous computing (Ubicomp) is a term coined in 1988 by Mark Weiser and defines the concept that computing power can be available everywhere, embedded in every object that users interact with.

Continuous improvements in integrated circuit (IC) manufacturing, as predicted by *Moore's Law* (Moore, 1998, 2006), enabled not only increased computing power but also the ability to fit more of that power in increasingly smaller form-factors.

Interestingly, if we take a step back in time and look at the history of computation from a broader scale, Moore's Law can be interpreted as an extension of a trend that had already started in the early 1900s. The fact is that computational power has been growing at an exponential rate for more than a century. Ray Kurzweil (2005) captured this trend in the following graph, which uses a logarithmic scale for the Y-axis.

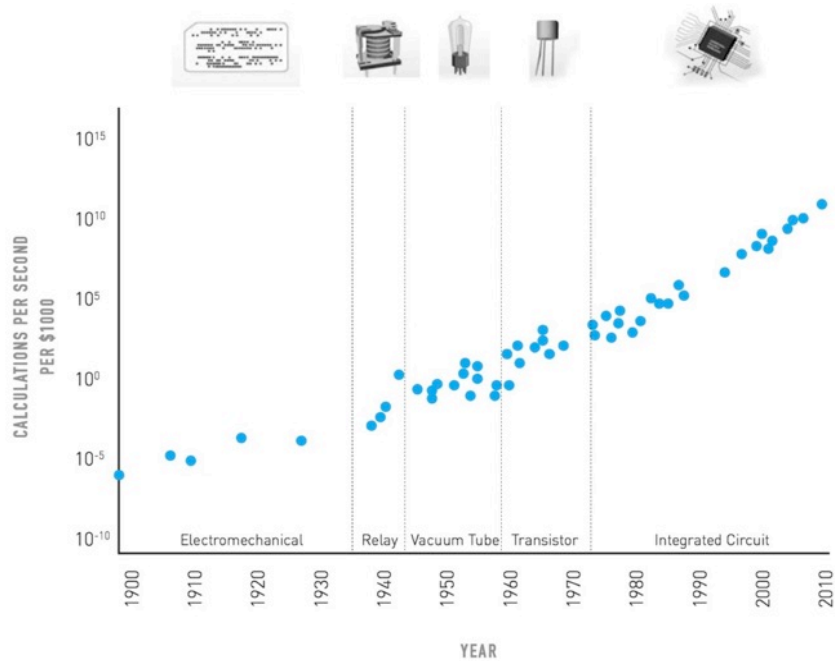

Source: (Kurzweil, 2005, p. 70)
Figure 1 - Exponential Growth of Computing for 110 Years
Note: Each dot is a computing machine (the most computationally powerful at the time)

Mainframe computers that traditionally occupied whole rooms gave way to the Personal Computer (PC) that could fit on a user's desk. Portable computers, tablet devices and smart-phones continued the trend, and allowed mobile computing. More recently smart watches and other wearable devices ensure that persons are seldom found without some form of computing power at their disposal at any given moment.

This trend towards smaller devices also had an obvious impact on the types of user interfaces, with increased simplicity happening as a by-product. Donald Norman, famous for coining the term "user-centered design" (1988), predicted in the late 90s that "invisible" information appliances were not only inevitable but a desirable alternative to the complexities of the PC (1998).

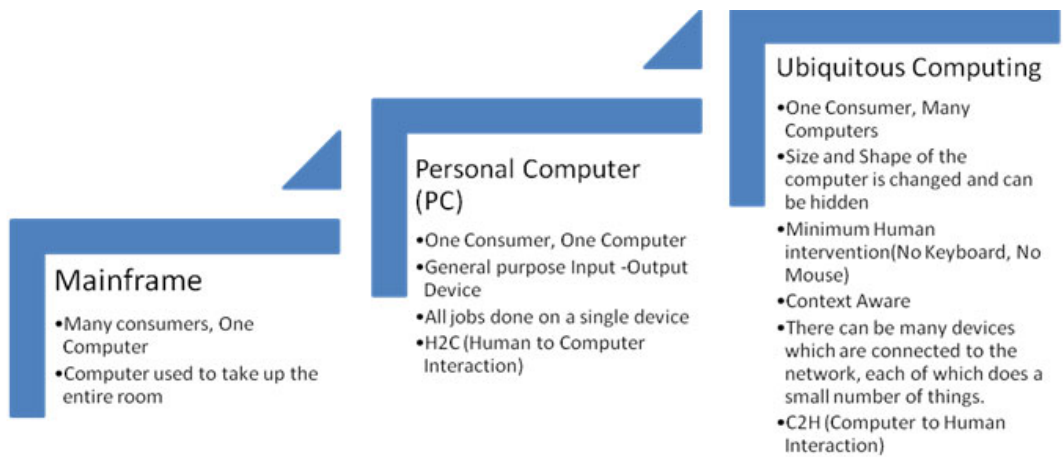

Source: ("Ubiquitous Computing – Living in a Smart World," 2004)
Figure 2 - Progression from Mainframe to Ubiquitous Computing

Most importantly, the users no longer interact with computers in the traditional sense. The interaction is increasingly with every-day items that happen to feature embedded computing capabilities.

### 2.2.2. THE INTERNET OF THINGS

Communication technologies have also played an important role in the trend towards always-connected devices. The creation of the Internet and advances in both high-bandwidth technologies like fibre optics, capable of transmitting data at multiple Gigabits per second, as well as wireless communication technologies, such as cellular networks (i.e. GSM), WiFi, Bluetooth, Near Field Communication (NFC), and ZigBee, amongst others, ensure that almost every device, no matter how small, can remain connected to the Internet and become a node in what is now popularly called the Internet-of-Things (IoT).

The potential growth for the IoT industry is significant and, according to John Chambers, CEO of Cisco, it could become a US$19 trillion market (Kharif, 2014).

2.2.2.1. **SMART PERSONAL DEVICES**

The embedding of computing in everyday appliances is also reflected by the growing trend of labelling these appliances as "smart". It is easy to forget that less than 20 years ago mobile phones were used primarily for making and receiving phone calls whilst on the go. Nowadays the voice calling function in smartphones is arguably one of the lesser-used ones, giving way to web browsing, instant messaging, photo and video recording, and usage of other apps, ranging from gaming to social networking.

Smartphones however, are only the beginning. Television sets are also becoming "smart TVs". Connected to the Internet, these devices can download various kinds of content from the cloud, and run a number of network-dependent applications on them, including voice and gesture detection, a controversial feature that will be further discussed later.

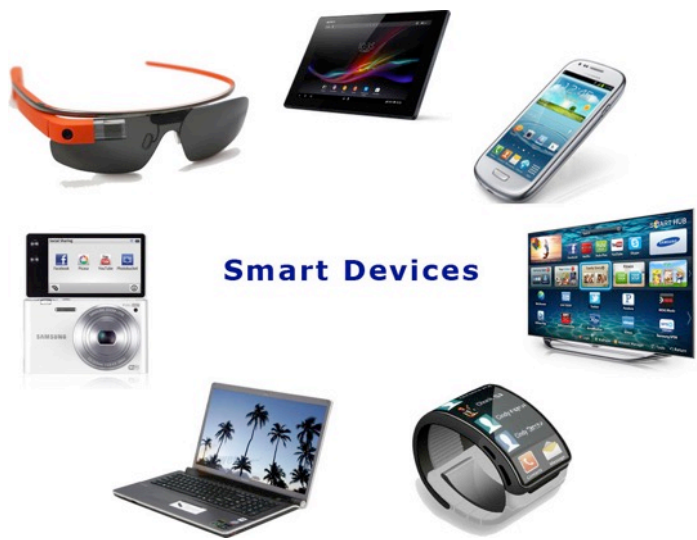

Source: (Cruze, 2014)
Figure 3 - A Range of Smart Devices

Several other types of every day items are becoming interconnected and available as smart devices on the market. Examples are the Ilumi[1], a smart light bulb that allows remote control via mobile phones, Nest[2] a smart thermostat that tracks and learns room occupancy patterns over time in order to save energy, the Luna[3] smart mattress that learns from user's sleeping habits (and potentially other activities) and that features a smart alarm and individual temperature control for each side of the bed, and the Aria smart weighing scale from Fitbit[4]. During meals users can also track what they eat with SmartPlate[5], how fast they eat it with the HAPIfork[6] smart fork, and the quantity and type of drink that they consume from the Vessyl[7] smart cup. After each meal the internet-connected smart toothbrush Kolibree[8] will record their user's brushing technique, providing insights into their tooth hygiene practices as well as turning the activity of brushing teeth into a game for the younger users. Even our home plants can benefit from smart pots like Click-and-Grow[9] that automatically provide the plant with the right amount of water and nutrients at the right time. Elsewhere in our increasingly smarter homes we will find smart toasters[10], smart fridges[11], smart garage doors[12], smart baby monitors[13], and smart yoga mats[14].

---

[1] http://ilumi.co
[2] https://nest.com
[3] http://lunasleep.com
[4] https://www.fitbit.com/uk/aria
[5] http://getsmartplate.com
[6] https://www.hapi.com/product/hapifork
[7] https://www.myvessyl.com
[8] http://www.kolibree.com/
[9] http://www.clickandgrow.com/pages/what-is-click-grow
[10] http://www.brevilleusa.com/smart-toaster.html
[11] http://www.samsung.com/us/appliances/refrigerators/RF28HMELBSR/AA
[12] https://garageio.com/
[13] http://www2.withings.com/us/en/products/baby/smart-baby-monitor
[14] https://smartmat.com/

Nearly every month the technology sections of crowdfunding websites such as Kickstarter[1] see new campaigns raising funds for the next household item that hasn't yet received the *smartening* treatment.

2.2.2.2. WEARABLE COMPUTING

Embedded computing and smart devices are also making their way into items that people wear. If the electronic wristwatch is itself not a novelty, having been introduced in the 1970s, more recent smartwatches are pushing the boundaries of connectedness in wrist wearable devices. Accessing basic computing functionality on the wrist is very convenient, including the ability to glance at the wrist to quickly check notifications that would otherwise involve removing a smartphone from a pocket. Both Apple, with the Apple Watch, and Google, with a range of Android Wear smartwatches manufactured by various partners, have recently pushed this technology to the hands of a fast growing user base.

Parents who want to keep track of the whereabouts of their young children are also increasingly purchasing wearable GPS tracking watches and bracelets, such as the Tinitell[2].

Between 2013 and 2015 Google experimented with an optical head-mounted display (OHMD) called Google Glass. This device has the capability to display basic information to the user on its OHMD and accepted input both via speech and a touchpad on the side of the device. It was also loaded with sensors such as accelerometer, gyroscope, magnetometer, ambient light sensor and proximity sensor. Arguably the most controversial feature is the head-mounted camera, which allows the user to take pictures or make video recordings with simple voice commands. As will be discussed in section 2.3.2, the camera feature pushed the threshold of what people were willing to accept in relation to privacy. If the Google Glass is being adopted for some industrial applications, it was clearly rejected by the consumer market.

---

[1] https://www.kickstarter.com
[2] http://www.tinitell.com/

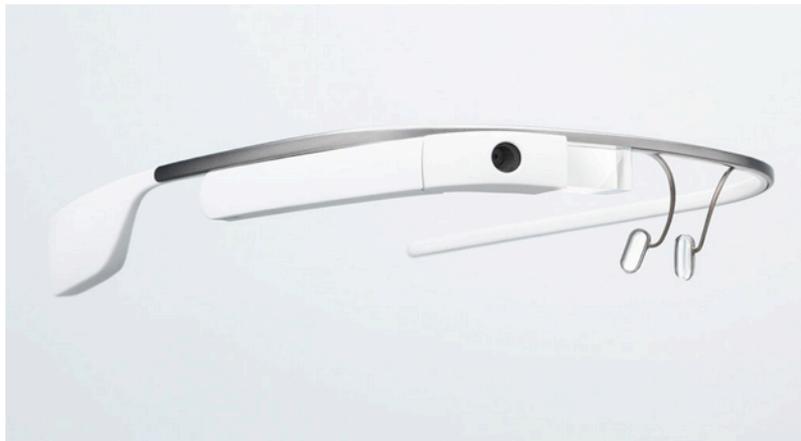

Source: (Reynolds, 2015)
Figure 4 - Google Glass

Having said that, several companies are working on wearable glasses technology for commercial use, and an especially interesting application is eye tracking. The ability to not only record a scene from the user's point of view (PoV) but to also track with accuracy the point in space where the user's eyes are focusing on has many applications, not least in usability and marketing research. Google filed for several patents in this area, hinting that the Google Glass would at some stage incorporate that feature (Calderone, 2014). SensoMotoric Instruments (SMI) is another company that already produces eye-tracking wearables, such as the Eye Tracking Glasses 2 Wireless (SMI ETG 2w) product.

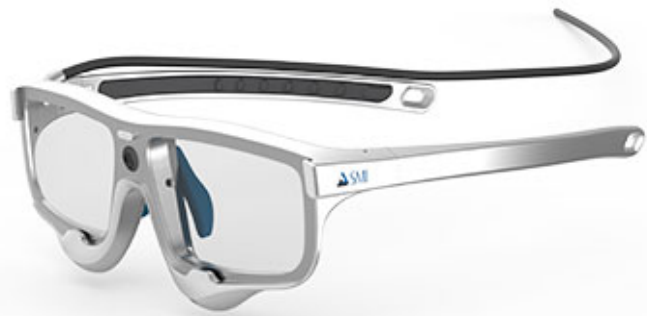

Source: (SMI, 2015)
Figure 5 - SMI Eye Tracking Glasses 2 Wireless (SMI ETG 2w)

Fitness and wellness related wearables have also been reaching the market recently, with a wide range of activity trackers available led by the popular Fitbit[1] and Jawbone[2] brands. Activity tracking technology is also finding its way into smartwatches and smartphones, with several Android device manufacturers integrating hardware sensors that efficiently replicate the features of dedicated fitness trackers directly in Android handsets and wearables. New Android APIs also allow app developers to efficiently access sensor data in a way that minimises battery consumption (Google, 2015).

Finally, in the area of security and authentication, Nymi[3] offers a wearable bracelet that uses biometrics, namely the user's heart rate footprint, to verify that the correct user in indeed present and adds another factor into any existing multi-factor authentication systems.

---

[1] http://www.fitbit.com/
[2] https://jawbone.com/
[3] http://www.getnymi.com/

2.2.2.3. **IMPLANTABLE AND INGESTIBLE COMPUTING**

In addition to embedded devices and sensors worn in clothing, technology has been already been adapted to be implanted in human bodies, typically for health reasons. Pacemakers and even artificial organs, such as hearts and lungs can provide life-saving temporary replacements until a valid organic organ becomes available through organ donation programs. Implantable continuous glucose sensors such as those produced by GlySens[1] and Senseonics[2] provide people with diabetes with real-time information on their smartphones.

Doctors will soon prescribe ingestible "smart pills". The first such device, that tracks whether patients with mental illness problems are taking their medicines on schedule, has already been approved by the US Food and Drug Administration (FDA) and is expected to go on sale towards the end of 2015 (Ward, 2015).

**2.2.3. LOCATION-BASED SYSTEMS AND GEOFENCING**

Location-based systems or services (LBS) date back to the early 1990s with the introduction of the indoors Active Badge System (Want, Hopper, Falcão, & Gibbons, 1992) which allowed the tracking of infrared-based devices worn by employees whilst in the office.

Since then we have seen a proliferation of LBS both in commercial and consumer settings, using location extracting technologies that range from satellite global positioning system (GPS), mobile network cell tower positioning, and WiFi access point tracking. For example, vehicle tracking is used in both contexts, with companies tracking their vehicle fleets and consumers cars featuring theft-tracking systems.

---

[1] http://glysens.com
[2] http://senseonics.com

Consumers are increasingly actively providing their location data by using social networks that allow the users to "check-in" to physical locations, like restaurants, cinemas, and other public locations. This is probably the least privacy invading LBS data acquisition method since the users are proactively and consciously sharing their location publicly, although as will be discussed later, the users are seldom aware of the potentially far-reaching implications of sharing such kind of personal data.

Other LBS systems that are popular with consumers include car navigation systems[1], discovery of nearby services[2], fitness activity mapping apps[3], family member tracking apps for those who want to know the whereabouts of their loved ones[4], and even location based massive multiplayer games[5].

A more recent method of location tracking that is gaining momentum in commercial contexts is called geofencing. It involves the use of low-power active beacons positioned in close proximity to create "virtual fences" that can be used to pinpoint with accuracy the location of consumers in commercial spaces like shopping malls (Namiot & Sneps-Sneppe, 2013) in order to gain insights into consumer behaviour patterns and also to provide them with relevant real-time information on their smartphones. Apple and Google are currently pushing two competing beacon communication standards, namely iBeacon and Eddystone. Two prominent beacon manufactures are NewAer[1] and Estimote[2].

---

[1] http://tomtom.com
[2] https://foursquare.com/
[3] http://www.mapmyfitness.com
[4] http://www.life360.com
[5] https://www.ingress.com

**2.2.4. CONTEXT**

To a large extent, the motivation for the current project stemmed from the author's determination to develop context aware applications whilst avoiding producing privacy invading technologies.

If the current project aims to produce a privacy friendly framework, it is undoubtedly to pave the way for future research into context, and therefore a review of context-related literature will help frame the future work.

Anind Dey provides the following definition of context:

*"Context is any information that can be used to characterize the situation of an entity. An entity is a person, place, or object that is considered relevant to the interaction between a user and an application, including the user and applications themselves"* (Dey, 2001).

Dey and Abowd go further in categorising context into tiers. Location, identity, time and activity are considered primary context tiers, which map to *who, what, what* and *when*, whereas other information such as phone numbers, addresses, list of friends, etc, are considered *secondary* tiers, since they can be extracted from various data sources given the primary tier. In other words, the primary tier context can be considered an index to the secondary tier context (Dey & Abowd, 1999).

---

[1] htttps://newaer.com/
[2] http://estimote.com/

## 2.2.4.1. CONTEXT-AWARE COMPUTING

Context-aware computing is a term coined by Schilit and Theimer (1994) and it defines a category of computer systems that can:

1. Sense the user's environment and context, and
2. Adapt to that context to better serve the user.

Context-aware computing has recently gained attention from major industry companies such as Google, Apple and Microsoft.

## 2.2.4.2. CONTEXT-AWARE PERSONAL ASSISTANTS

Google made significant progress into context-aware computing with their Google Now personal assistant. Available in both Android and iOS, it uses a speech recognition interface along with a natural language processing engine to make recommendations and answer questions, whilst making use of the user's context to filter the most relevant information. It can even pro-actively nudge the user with relevant information at the right time, without the user requesting it.

Apple, which already featured its Siri personal assistant on iOS devices, also focused on making it more proactive like Google Now, with the introduction of the *Proactive Assistant* update to Siri (Lardinois, 2015).

Microsoft similarly added contextual-awareness to its own personal assistant, Cortana, by making use of structured data in e-mail messages (Microsoft, 2014).

## 2.2.5. QUANTIFIED-SELF

We are witnessing a boom in the adoption of wearables for self-tracking. The now called quantified-self movement (G. Wolf, 2014) is growing at a considerable rate. ABI Research estimated that sports and activity trackers will exceed 485 million annually shipped units by 2018 (ABI Research, 2013).

The potential for fitness and health tracking is especially promising. England's Rugby Sevens team employ a system developed by StatSports[1], which tracks every player during training. The system's wearable sensors feature GPS, accelerometer, magnetometer, and gyroscope, to track in real time the smallest detail in the player's movements. The team's head physical performance coach, Brett Davinson, claims that the data is so sensitive and accurate that they can predict when a player is susceptible to injury or illness 24 hours before the player himself is conscious of any symptoms (Fong, 2013). Similarly, Dr Eric Topol is leading the way in prescribing mobile apps for self-tracking as an alternative to traditional medicines for treating or managing chronic diseases. According to Dr. Topol, heart rate, blood glucose and even air quality and pollen sensors, can go a long way to predict and possibly prevent life-threatening crisis (Fong, 2013).

## 2.3. THREATS TO PRIVACY

If on the one hand ubiquitous computing and context-aware computing have the potential to revolutionise the way we interact with technology and provide significant benefits to individuals and society, on the other hand the risks in terms of privacy are equally substantial.

The problem stems from the fact that enterprises are increasingly turning to consumer tracking and consumer data as a business model.

---

[1] http://statsports.com/

## 2.3.1. KNOW YOUR CUSTOMER

"Know your customer" (KYC) is a well-established tenet of marketing, with consumer profiling being the foundation upon which other marketing activities depend on (Brassington & Petitt, 2006). This recording and classification of consumer behaviour relies on multiple methods of consumer tracking, including credit card transactions, loyalty card usage, online tracking, web searches, and more recently face recognition (N. Wolf, 2012) amongst others. Furthermore, large online service operators such as Google and Facebook, take advantage of the large amounts of personal consumer data stored on their data silos to engage in behavioural analysis (Elmer, 2004) for the purpose of selling targeted advertising (Rushe, 2013) (Moriarty, Mitchell, & Wells, 2009).

Nowadays most major retail stores track their customers through credit card numbers and analyse their shopping behaviour for consumer insights (Duhigg, 2012). One such chain, the US based Target, famously calculated that a teenage girl was pregnant before her own family found out (Hill, 2012).

## 2.3.2. THE COST OF FREE AND THE END OF PRIVACY

The World Wide Web has evolved significantly since its original proposal by Sir Tim Berners-Lee (Berners-Lee, 1990). Even though it is nowadays highly commercialised, most services remain free of charge for the users.

Traditionally the business model that sustains free online services is advertising, but if on the one hand advertising per-se does not entail privacy issues, on the other hand services are increasingly turning to consumer profiling techniques in order to improve the relevance of those adverts. In effect, "free" online services are not really free. Users are in fact paying for those services with their personal data. When Google launched their free web mail service, Gmail, it sparked a privacy row (Rushe, 2013) (BBC, 2004). Privacy conscious users were concerned about the fact that Google was snooping on the content of emails in order to serve more targeted advertising. However Gmail remains a very popular email service, and paved the way for others to follow suit in conducting similar privacy invading practices. From the point of view of the consumer, the convenience of accessing free services seems to outweigh the privacy issues inherent with it. Trust in well-known web service providers may also play a factor in this complicity. However trusting personal data to commercial entities also carries the risk that those entities may one day be sold or go bankrupt. In such cases, the customer data is considered a commodity that is sold to the highest bidder, along side the other assets of the company (Singer & Merrill, 2015).

### 2.3.3. IOT TRACKING

If offline and web-based tracking carry privacy risks, the introduction of IoT raises these issues to unprecedented heights.

Manufacturers of smart devices are increasingly waking up to the potential value of consumer data, and as such are engineering their IoT devices to automatically upload the user's data to their own cloud. Smart TV manufacturers LG were investigated by the UK's information commissioner when it was found that their smart TVs were sending information about viewing habits as well as files stored on attached USB disks back to Korean company (Arthur, 2013). More recently Samsung were in the news for the wrong reasons, when it was reported that their voice controlled smart TVs were sending recordings and transcripts of their user's private conversations back to the manufacturer, in order to improve their voice recognition software (Bigelow, 2015). This worrying development invoked memories of George Orwell's *Nineteen Eighty-Four* novel and its famous Telescreen appliance, which acted as both a television and a CCTV camera (Orwell, 1949).

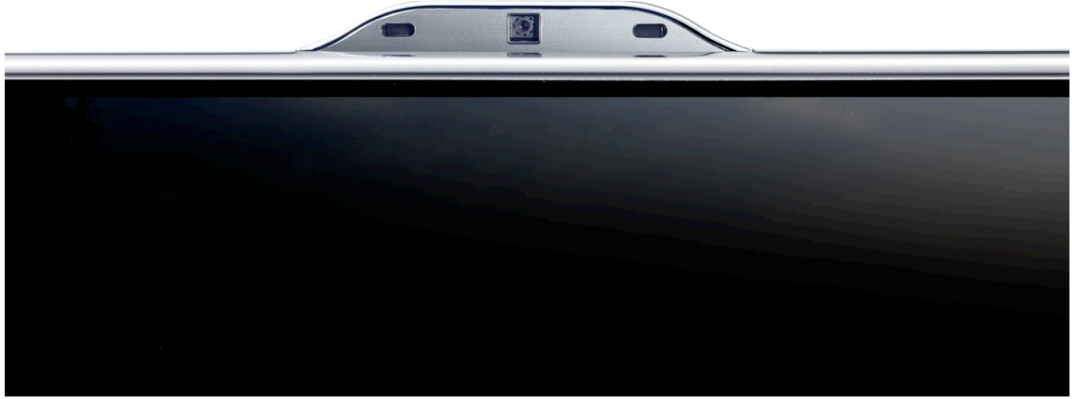

Figure 6 - Detail of a Samsung Smart TV

Google Glass, in another example, quickly raised privacy concerns from people who were reluctant to be recorded in social and private contexts. The device was famously banned from bars, restaurants and several other types of establishments, and the derogatory term "glasshole" was popularly adopted to define users of the device (Schuster, 2014). Google eventually discontinued the project with privacy issues being commonly referred to when justifying its failure (Bilton, 2015).

At this stage it is worth to contrast the differences in consumer resistance towards Google Glass versus the increasingly pervasive CCTV technology. Although both technologies record people, the more intimate and visible nature of Google Glass, plus the fact that it also recorded sound, which is seldom the fact with CCTV, meant that people were not capable of accepting the former. It is therefore important to point out that, even in the current context of overall diminished privacy, there are still some privacy-related thresholds that people will simply not allow to be violated.

**2.3.4. BIG DATA**

The large amounts of consumer data gathered by enterprises, and subsequence analysis for gaining consumer insights has led to the coining of the term big data. Although big data is itself not a new paradigm, the recent attention gathered around it has created a thriving industry of companies that specialise in everything from storing to data mining. One of the problems created with storing large amounts of data in centralised data silos is that it attracts the attention of tech-savvy criminals who hope to profit from stealing and selling on the data.

**2.3.5. SECURITY ISSUES**

When enterprises are not willingly violating the trust of the consumers through explicit privacy invading practices, they often do so accidentally due to poor security standards around the storage of the consumer data.

Even governments, who are expected to keep to the highest standards in information security, regularly leak large amounts of private information. For example, the US Federal officers database was recently compromised, leaking between 14 and 21 million personal records and putting each and every one of those persons at risk of identity theft or worst (Perez & Prokupecz, 2015).

A couple of attack vectors will be explored below.

### 2.3.5.1. THERMAL ATTACK ON AIRGAPPED COMPUTERS

A new attack vector using thermal radiation was recently demonstrated. By controlling the throttling of CPU and GPUs a machine can create heat patterns that can be detected by a nearby system, even if both systems are air gapped (Zetter, 2015) (Guri, Monitz, Mirski, & Elovici, 2015). The attack does require both systems to be compromised in advance, in order to collude in the data transfer, in close proximity, and with built-in temperature sensors. Most modern CPUs come with such built-in sensors.

### 2.3.5.2. TEMPEST ATTACKS

TEMPEST attacks involve the detection of minute electromagnetic fluctuations around the system under attack, in order to extract valuable information such as a private key used in encryption. Normally these kinds of attacks require specialised hardware and physical access to the target machine, but there are other situations in which two machines in close proximity can communicate via electromagnetic or radio waves.

### 2.3.5.3. SUCCESSFUL ATTACKS ON PEER-TO-PEER NETWORKS

If on the one hand centralized silos with large amounts of personal data act as honey pots and attract the attention of malicious hackers, on the other hand having a decentralized network of nodes is not by itself a guarantee of resilience against attacks, especially if any aspect of the network relies on a centrally accessible services or directories of nodes that can be easily targeted by malicious hackers. Such services, such as Dynamic DNS (DDNS) or Remote Access Services, offer convenient ways for users to access their home devices from anywhere, however the author believes these services restore the undesirable honeypot aspect of the network. Taking the Synology NAS device as an example, the author found anecdotal evidence that links the setting up of a device with a Dynamic DNS service with a significant high number of unauthorized login attempts to said device. Furthermore, when a Synology vulnerability is exploited, it is invariably the case that numerous users are affected in a short period of time, showing that the attackers know exactly which IP addresses to target when launching the attack. Synology has gone so far as to disable the remote access services from nodes that are found to be running exploitable versions of their operating system. Whilst this is a sensible procedure for mitigating known attack vectors, the fact is that users of such services are still vulnerable to zero-day attacks.

Additional attack vectors rely on homogeneous default configurations amongst the peers, such as using a well-known public port number. Again using Synology as an example, when a vulnerability was found in the operating system, network security firms detected a sharp rise in Internet port scanning specific to the default port used by Synology NAS devices, in this case port 5000, as shown in Figure 7. This particular event happened in February 2014, when news reported that Synology devices were falling victim to a remote attack that turned the devices into cryptocurrency miners, netting the attackers an estimated USD $600,000 (Litke, 2014).

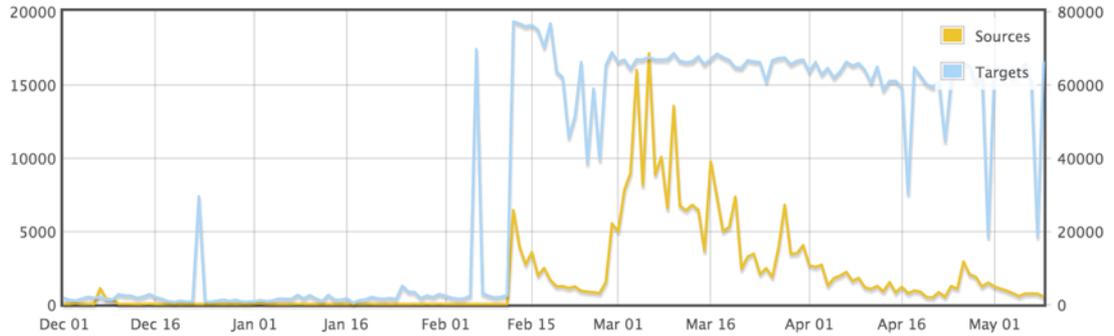
Source: (Litke, 2014)
Figure 7 – Scan activity for port 5000, Feb 1 2013 through May 9 2014

## 2.4. EXISTING PRIVACY PROTECTION EFFORTS

Some efforts have been made to protect consumer privacy, and the following sections will highlight some of them.

### 2.4.1. REGULATIONS

Regulation aims to restrict what data enterprises can collect, for what purposes and for how long it can be stored. However regulations create friction that the enterprises will attempt to work around. As long as the incentive is to collect personal data, enterprises will always find a way to achieve their goals whilst remaining on the right side of the law.

Some industries do in fact try to self-regulate in this regard, through non-profit organisations such as the Network Advertising Initiative (NAI)[1], and the Digital Advertising Alliance (DAA) Self-Regulatory Program[2] but one must question the effectiveness of these efforts coming from an industry that is increasingly more dependent on mining consumer data.

---

[1] http://www.networkadvertising.org/
[2] http://www.aboutads.info/

## 2.4.2. PERSONAL DATA STORES

With the rise of the web and the *cloud* in recent years, and subsequent control over our data by enterprises, there is a growing feeling that we are entering an age of "digital feudalism" (CGCS, 2012) and as a result a number of projects have been started which aim to provide user-controlled alternatives to the *cloud*. Existing projects will be compared against the proposed project via a set of criteria, namely whether they provide a:

- Personal Server (PS) device that hosts the personal data and digital-self:
    - Hardware
    - Software
    - Behavioral Data Analysis
- Self-tracking software (STS) that sends data exclusively the PS;
- Privacy-Protecting (PPAPI) to the digital-self.

The personal server/cloud projects can be divided into 2 categories, a) software and b) hardware solutions:

a) Software projects need to be installed on commodity hardware, such as a personal computer, a small plug computer or a Network Attached Storage (NAS) device. This does require considerable know-how by the user, as well as resulting in a system that is poorly integrated, and potentially insecure.

Software projects include, OwnCloud[1], CozyCloud[2], FreedomBox[3], Sneer[4], and Unhosted[5].

b) Hardware solutions comprise of both software and hardware integrated into a product, which normally provides a better user experience. These include the TonidoPlug[6], SpaceMonkey[7], and Transporter[8]. These projects currently aim to provide self-hosted alternatives to *cloud*-based file storage services such as Dropbox[9].

Table 1 summarizes how these projects fare against the above-mentioned criteria. As can be seen, none of them apply behavioral data analysis to the user's data, nor do they provide a privacy-protecting API for enterprises.

|  | PS | | | STS | PPAPI |
| --- | --- | --- | --- | --- | --- |
|  | **Hardware** | **Software** | **Data Analysis** | | |
| CozyCloud | No | Yes | No | No | No |
| FreedomBox | No | Yes | No | No | No |
| OwnCloud | No | Yes | No | No | No |
| Unhosted | No | Yes | No | No | No |
| TonidoPlug | Yes | Yes | No | No | No |
| SpaceMonkey | Yes | Yes | No | No | No |
| Transporter | Yes | Yes | No | No | No |

Table 1 - Existing projects do not fulfill required criteria

---

[1] http://owncloud.org
[2] https://www.cozycloud.cc
[3] http://freedomboxfoundation.org
[4] https://sneer.me
[5] http://unhosted.org
[6] http://www.tonidoplug.com
[7] http://www.spacemonkey.com
[8] http://filetransporterstore.com
[9] http://dropbox.com

It should be mentioned that a large number of mobile apps do exist for self-tracking (STS), such as *Expereal*[1] for mood tracking, *Sleep as Android*[2] for sleep tracking and *RunKeeper*[3] for tracking fitness activities, however none of them integrate with a PS, and those that synchronize data with an external system choose to do it on servers controlled by the app developer, outside the control of the consumer. These apps obviously follow the traditional *cloud* approach that the proposed project aims to replace, and hence were omitted from this analysis.

## 2.5. SUMMARY

In summary, even though there are a large number of projects that do attempt to address the diminishing power of consumers in the traditional *cloud*, as well as provide them with increased privacy, none address the marketing needs of enterprises, and as a result these enterprises will continue to engage in privacy invading practices in order to achieve their goals.

Grokya will attempt to fill this gap by creating an optimum solution that simultaneously satisfies the needs of both consumers and enterprises.

---

[1] http://expereal.com
[2] http://sites.google.com/site/sleepasandroid
[3] http://runkeeper.com

## 3. SYSTEM DESIGN AND SPECIFICATION

*"If I had asked people what they wanted,
they would have said faster horses"* – Henry Ford

### 3.1. METHODOLOGICAL APPROACH

Similarly to the Henry Ford quotation that opened this chapter, Steve Jobs once famously said "*in the end, for something this complicated, it's really hard to design products by focus groups. A lot of times, people don't know what they want until you show it to them*" (Reinhardt, 1998).

This project follows Ford's and Job's philosophy by applying a Genius Design approach (Saffer, 2009), which relies primarily on the individual vision, inspiration, and skill of the designer, whose product is subsequently validated by the users. This approach was chosen because, as explained in the previous chapter, the overall concept of the project presents a significant departure from previous concepts and there is no prior art that the users can relate to.

### 3.2. GROKYA GUIDING PRINCIPLES

Before designing the proposed solution a few guiding principles were established in order to ensure all future decisions conform to the designer's vision. When making design and implementation decisions, those decisions should be influenced by these principles as much as possible. These principles will be referred to as Grokya Guiding Principles (GGP):

1. Technology should always adapt to the user, not the other way around;

2. An invisible and unobtrusive system is always better than one that interrupts the user or requires constant manual feedback;

3. A transparent and trustless system is always better than one that requires trust on third parties;

4. Don't make the user think, use sensible defaults as much as possible, and ideally learn the desirable defaults from the user's behaviour;

5. Any system or any person who contributes to the generation of value (utility or economic) should receive a fair reward for their contribution;

6. Sharing of data cannot be undone. If privacy though data protection is the goal, then adopt a *share-nothing* strategy as much as possible. Always ask *what do you want to do with that data,* and share the higher-level desired solution instead of the raw data.

Following sections may make a reference to a particular GGP by use of its number. For example, GGP#4 is a reference to the "don't make the user think" principle.

### 3.3. OVERALL ARCHITECTURE

The main purpose of this project is to present a privacy-preserving alternative to traditional big data analytics. The plan is to achieve this by introducing a consumer owned personal server, herein called LifeServer (LS), as a viable personal data storage alternative to the traditional cloud. The main distinctions from existing consumer personal data store (PDS) solutions are:

- Focus on implicit data acquisition from IoT devices for contextual awareness;

- A trusted hardware platform, the LS, where private data can be securely stored;

- An API that allows for enterprises to conduct high-level analytics and marketing activities, such as behavioural advertising, without direct access to personal data.

**3.3.1. CENTRALIZED VS DECENTRALISED VS DISTRIBUTED**

One of the main architectural questions was regarding the decentralisation of the platform.

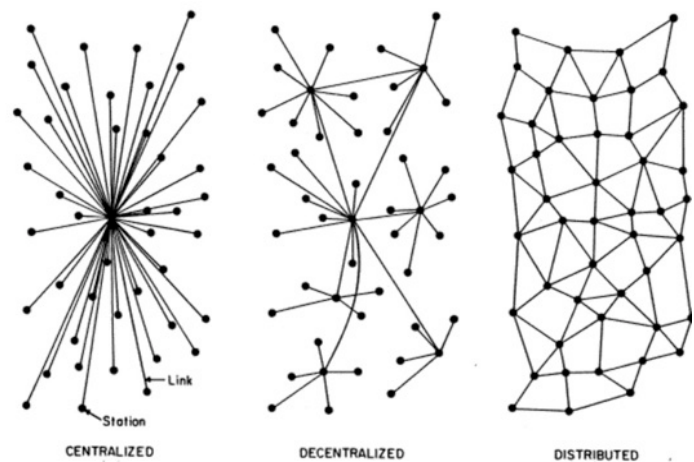

Source: (Baran, 1964)
Figure 8 - Centralised vs Decentralised vs Distributed

Centralized infrastructure represents a single-point of failure in terms of security. If a security breach occurs in a centralised architecture, then the whole system has failed and the all the data is compromised, or at least that data which is not encrypted. Central repositories of data also act as "honey pots", attracting the focused attention of malicious hackers who benefit the most from gaining access to a potential gold mine in terms of valuable data. Finally, a centralised architecture entails a central authority in charge of the infrastructure, on which the users must trust. This last point clearly goes against GGP #3.

It could be argued that even decentralised, or federated systems, also rely on trust on third parties, although to a much lesser extent that fully centralised systems. For this reason, the proposed project will aim for a fully distributed architecture, minimising the need for trust on any third parties.

### 3.3.2. DISTRIBUTED BIG DATA

Looking at the big picture in terms of big data, the goal is to allow consumers to reclaim their personal data and to keep it under their own exclusive control, by offering them the consumer infrastructure to do so with minimum effort. These consumer personal servers should also find and connect to other consumer's servers and create a p2p network that allows for collaborative data mining and analytics, without leaking personally identifying information. This p2p network of anonymised data analytics will then allow enterprises to extract high-level insights and perform other marketing related activities.

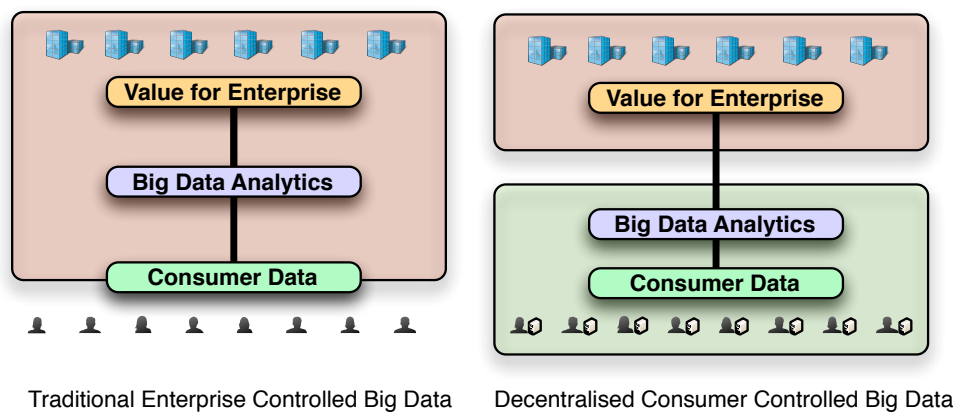

Traditional Enterprise Controlled Big Data    Decentralised Consumer Controlled Big Data

Figure 9 - Big Data: Centralised vs Decentralised

In the above figure, with enterprises on top and consumers at the bottom, the red shaded area denotes the enterprise domain, and the green the consumer domain. In the traditional big data model the moment that consumer data is generated it is stored in enterprise controlled data centres giving the enterprises full stewardship of the data with the resulting implications for consumer privacy. The proposed approach, on the right side of the figure, creates a consumer domain by providing consumers with their own infrastructure, solving the issue of privacy whilst still providing enterprises with their ultimate goal of extracting value from big data analytics.

### 3.3.3. VALUE DISTRIBUTION

Another important component of this project is to provide economic incentives to all the stakeholders who are already engaged in IoT, big data, and other potential privacy-invading practices to switch to the new proposed paradigm.

The reasoning here is that enterprises are not really interested in invading people's privacy. Their main goal is to maximise profitability, and it so happens that gathering and analysing large amounts of personal consumer data is currently the most effective way to achieve those goals.

Due to the distributed nature of this project a new protocol will be proposed: Value Distribution Protocol (VDP). This new protocol addresses the distribution of economic value amongst a large number of related stakeholders who contributed to the generation of that value using the Grokya platform.

### 3.4. API OVERVIEW

The following figure depicts a high-level view of the 3 main APIs (drawn in purple boxes) supported by the LS, and their role in the interactions with the IoT devices and the enterprises.

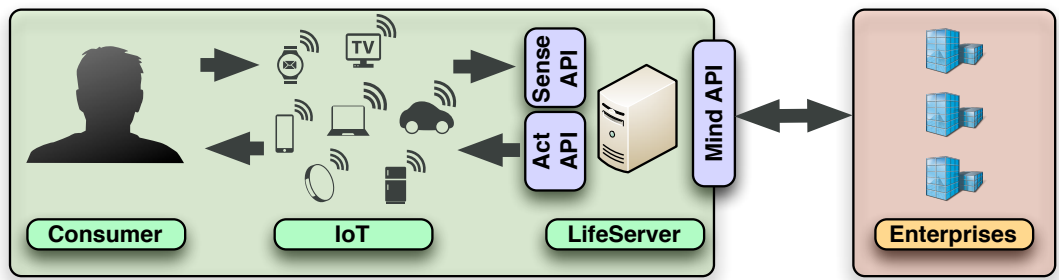

Figure 10 – API overview

The 3 APIs are:

- **Sense API**: a protocol that allows external IoT sensors, devices and apps to contribute personal data to the user's LS;
- **Act API**: a protocol that allows the LS to control and customise external IoT apps and devices, enabling the personalisation of user experience and digital content presented to the user;
- **Mind API**: a protocol and query language that allows third parties to gain insights from the user without being granted direct access to the user's personal data.

As clearly shown in Figure 10 above, the LS is not designed to interact directly with the user. Instead its main purpose is to interact with the user's IoT devices, in order to learn from user behaviour and customise the user experience through those same devices.

The APIs will be explained in more detail in the following sections.

### 3.4.1. SENSE API

The Sense API is the main personal data input channel of the LS.

All IoT devices and apps that the user interacts over the course of time should make use of the Sense API to contribute all the user-specific data that was extracted from those interactions.

This data will then be stored and made available for future data mining activities performed by the LS.

Every IoT device or app that submits personal data via the Sense API is called a *Data Source*. The API allows for requests to optionally include meta-data that links a particular portion of data to a Data Source. This meta-data will allow the LS to share some of the economic revenue that resulted from analyzing that data on behalf of enterprises.

### 3.4.2. ACT API

The Act API allows the LS to take control of external IoT devices and apps in pre-defined ways so as to allow the customization of the user experience.

The external devices and apps must expose a set of controls or configuration options that the LS will control during the user's interaction with said devices. For example a smart light bulb can expose the necessary controls that allow the LS to control the light according to the user's context. A modern car with electronically adjusted seats and mirrors can provide the LS with the controls that allow a quick re-configuration to the user's preferred or last known setup. This last use case also demonstrates how the car could have previously made use of the Sense API to provide the LS with any manual modifications that the user might have made.

The usefulness of the Act API is that, by keeping the user preferences stored in the user's LS, the user can effectively carry his personalized experience across a range of devices (such as cars), including those he or she never interacted with before.

### 3.4.3. MIND API

The Mind API allows third parties to interact with the user's LS in order to extract useful insights from the user and conduct high-level marketing activities without having direct access to the user's personal data.

Using the Mind API, marketing activities such as behavioural and targeted advertising will no longer require that third parties have access to the user's personal data. Instead the enterprises push the advertising meta-data to the LS, which in turn internally selects any ads that are of high relevance and interest to the user and displays them where and when appropriate via the Act API.

A Mind Query Language (MQL) should also allow for anonymised queries run over a large number of LS nodes, effectively providing statistical data from the network without identifying individuals.

### 3.5. VALUE DISTRIBUTION PROTOCOL (VDP)

Given the distributed nature of the Grokya platform, with each user controlling their own LS, the problem of fairly rewarding stakeholders that contributed to value being generated by the platform is non-trivial. This section will describe a solution to this problem, the Value Distribution Protocol (VDP). Even though VDP started as a small component of the Grokya project, it has generated enough interest as a stand-alone project that it was chosen to be presented at the Hackcoin blockchain hackathon in Hong Kong, winning the first prize at the event (Weese, 2015). The protocol is now developed independently of the Grokya project in it's own Github repository[1] and the development of reference implementation is underway under project name Subatoshi[2].

### 3.5.1. TRADITIONAL ADVERTISING REVENUE SHARING

In a centralised platform, such as the typical advertising networks, the process of assigning value is straightforward. Advertisers pay the network for adverts to be displayed, and the network shares some of that revenue with the content-providers that carry the ads to the consumers.

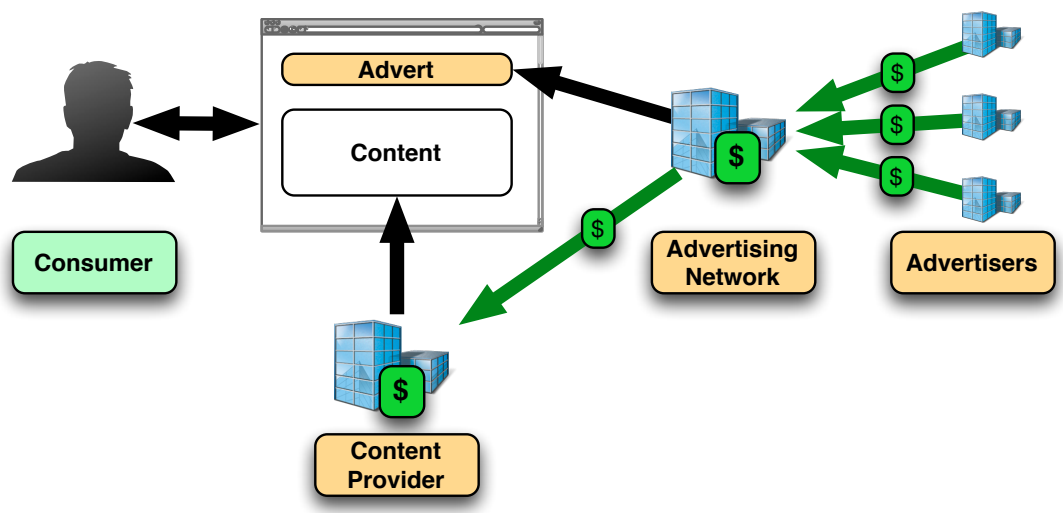

Figure 11 - Traditional Advertising Revenue Sharing

What this simple revenue sharing model diagram fails to capture is the trading of personal consumer data that increasingly takes place behind the scenes, as was discussed in the previous chapter. Figure 12 shows an example of data and economic value flows, with an advertising network as the final data silo.

---

[1] https://github.com/ktorn/vdp
[2] http://subsatoshi.org

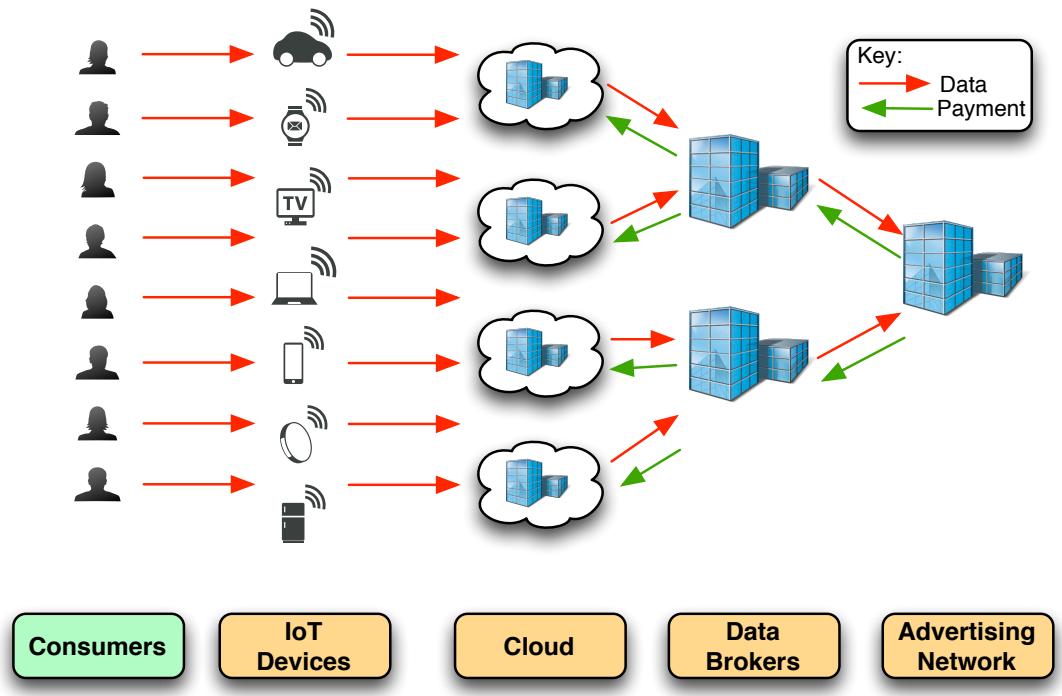

Figure 12 - Traditional Personal Data Brokerage Example

A core concept in this proposal is that every entity that contributes to the new user-controlled platform should be fairly rewarded (GGP#5). This reward should provide enough of an economic incentive so that those same entities can discontinue their participation in the traditional privacy invading data brokerage activities.

### 3.5.2. REWARDING DATA SOURCES

Examples of such entities are the IoT device manufacturers and smart phone app developers. The idea is that when a device or app contributes data to the LS, and if that data is later used in a transaction that generates economic value, then the device manufacturer or app developer should receive a small payment as a result of their contribution to making the transaction a success.

This means that every piece of data sent to the LS via the Sense API should be tagged with some kind of source identification metadata that uniquely identifies the source of the data. As will be explained below, this identification can be quite complex, since data can exist in various states of processing by different processes and aggregated by various sources. The proposed protocol needs to handle such complexities.

### 3.5.3. THE VDP PROTOCOL

The basic building block of VDP is its configuration file:

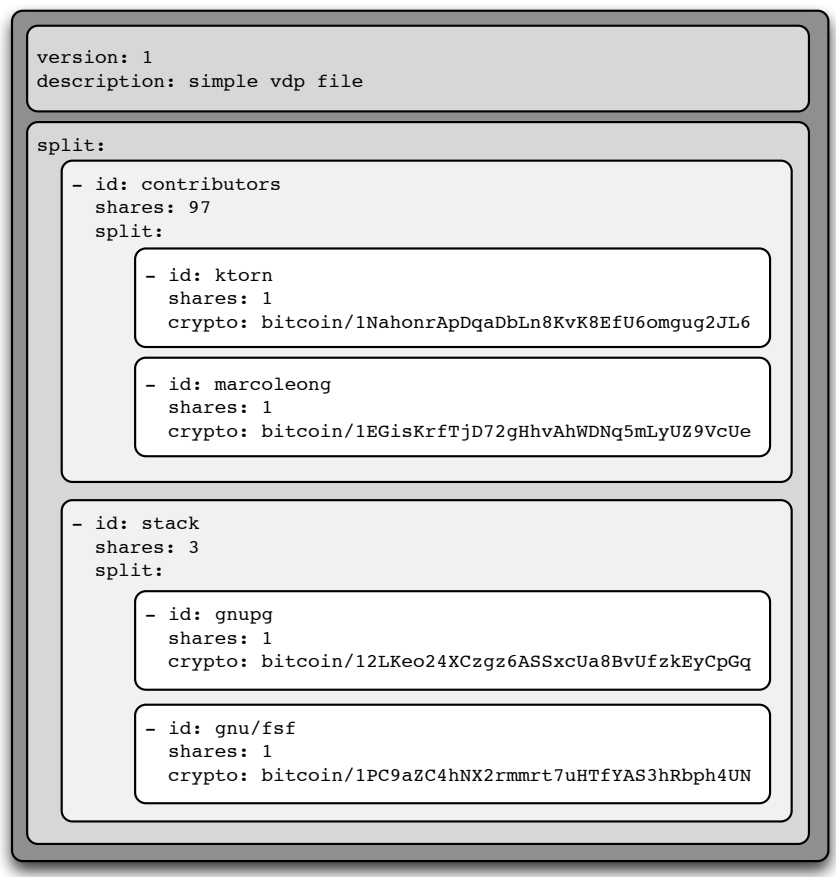

Source: (Farinha, 2015)

Figure 13 - VDP basic configuration syntax

A closer look at this example VDP configuration reveals 7 types of keywords:

- **version**: the protocol version, currently at version 1;
- **description**: a basic description of this configuration file or snippet;
- **split**: takes as input the value to be distributed, and splits it amongst the child nodes, according to their relative shares;
- **id**: identifies a node within a branch, which can be stakeholder, or another sub-branch. The id needs to be unique only within the scope of that branch;
- **shares**: the shares of a node. The shares are relative to the other sibling nodes' shares within that branch. For example, in Figure 13, under the *contributors* node, the sibling sub-nodes *ktorn* and *marcoleong* each have 1 share, which means they split 50/50 of whatever value flows down to that branch, which in turn consists of 97% of the total value attributed by that VDP configuration.
- **crypto**: a cryptocurrency address where the value can be paid into directly;
- **bitcoin**: indicates that the cryptocurrency address is a bitcoin address, with other cryptocurrencies also a possibility.

The power of VDP comes from its simple design, and also a key 8[th] keyword that is missing in the above example:

- **url**: a node can have it's configuration in an external file identified by a URL, either locally or in a remote server.

This means that a VDP configuration itself can be decentralized, with each VDP file hyperlinking to other VDP files, as exemplified in the following figure.

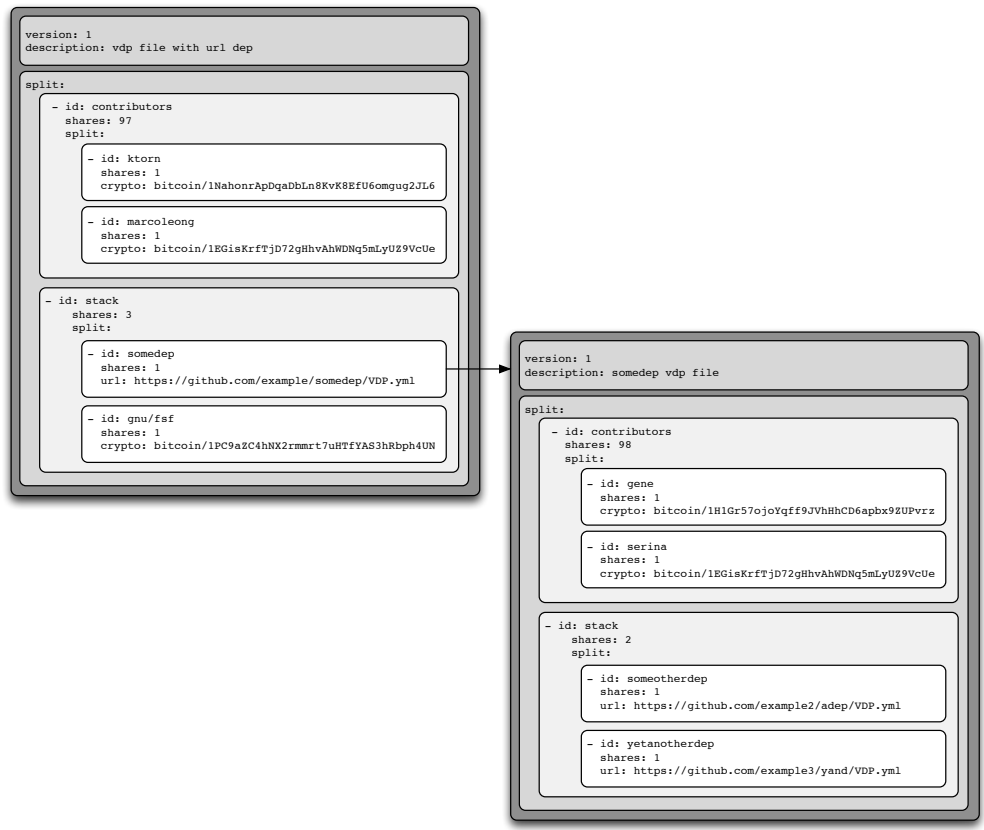

Source: (Farinha, 2015)
Figure 14 - Hyperlinked VDP files

### 3.5.4. VALUE COMPUTATION ENGINE

With the basic VDP protocol in place we can see how it plays an important part in a typical use case where an enterprise is willing to pay for a particular actionable insight.

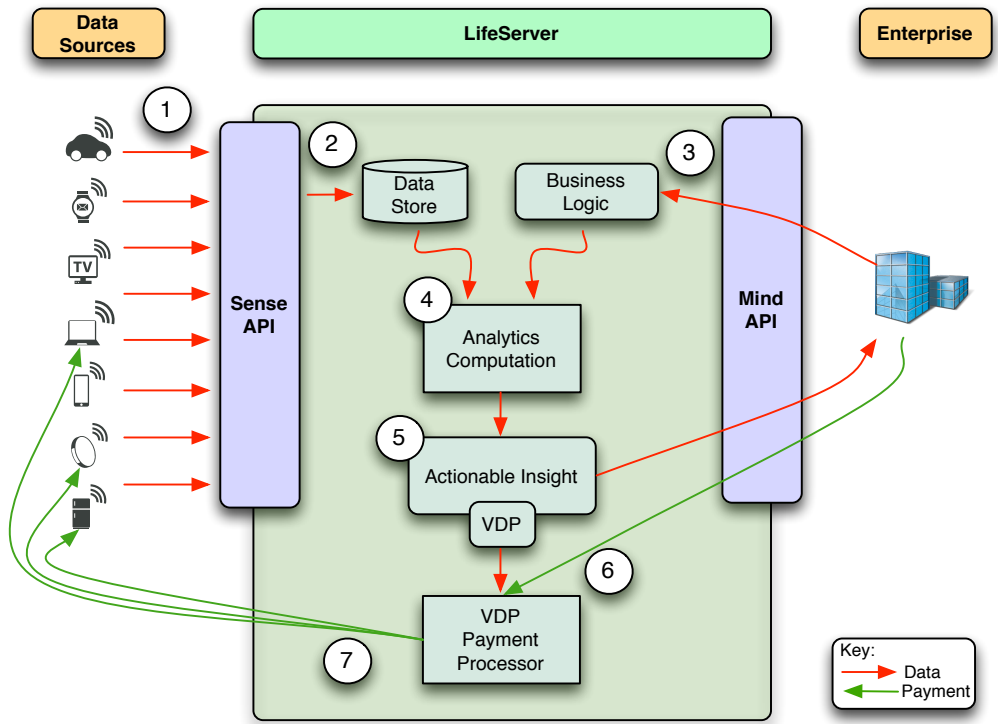

Figure 15 - Example of Value Distribution with VDP

In this example the following steps (indicated with numbers in Figure 15) take place:

1. The user's IoT devices and/or apps contribute user data to the LS via the Sense API. This data should include the device manufacturer's or app developers VDP payee instructions as meta-data;

2. The LS stores this data for later analysis. The stored data contains the data source meta-data;

3. The LS receives, via the Mind API, a request from an enterprise. This request includes the business logic that enterprise whishes to run against the locally stored user data;

4. The LS takes the business logic runs it against the stored user data. If the LS contains all the necessary data to perform the computation then the computation succeeds;

5. If the computation in step 4 completes successfully, the output will be both the actual computation result, i.e. an actionable insight that is valuable to the enterprise that requested it, and a VDP file that includes all the data sources that contributed to the data involved in the computation.

6. A transaction is made via the Mind API between the enterprise and the LS. The enterprise pays a fee in return for the actionable insight.

7. The LS takes the fee paid by the enterprise and the VDP file outputted in step 5 and processes the VDP payments, making a series of micropayments towards the manufacturers and developers of the data sources that contributed the user data used in step 4.

## 3.6. LIFESERVER HARDWARE

This section details some of the decisions taken with regards to the LifeServer hardware. Most decisions will take into account the GGPs outlined in page 40.

### 3.6.1. HOSTING: SELF VS CLOUD

With regards to the hosting of the user's personal data, given the privacy implications of the traditional cloud-based big data paradigm one would naturally be inclined to choose self-hosting as a viable alternative. However the pros and cons of this decision should be explored in more depth.

According to GGPs #1, #2, and #4, which are mostly concerned with technology that gets out of the way of the user, a cloud solution would actually provide the best option, since it doesn't require any installation or maintenance by the user. However as discussed earlier, trusting a third party with personal data clearly goes against GGPs #3 and #6.

The challenge then is to adopt a solution that does not violate these key GGPs and still achieves a relative ease of use and easy to maintain system.

It is important to note that the issue with hosting data on the cloud is only an issue as long as the cloud operator has full access to that data. This is not always necessarily the case, since the data can be encrypted on the user side before being uploaded to the cloud. In such a case, the cloud provider will only be storing a blob of unreadable binary data, which does not raise any privacy issues. However this also means the cloud operator will not be able to perform any processing of that data to the benefit of the user, so the user would still have to rely on a device that would download the data from the cloud, decrypt it, and process it. Fully homomorphic encryption (Gentry, 2009) is a field of research that aims to allow computation directly on encrypted data, and is making progress towards that goal, however at present it is extremely inefficient for any practical application (Yegulalp, 2014).

For this reason the decision is to rely on self-hosting with optional cloud-based backups of encrypted data.

### 3.6.2. DEDICATED VS SHARED HARDWARE

The easiest way to get started with LS would be to just install the LS software on the user's existing hardware, such as their smartphone or PC. However these are devices that the user interacts with directly, and which allow the installation of arbitrary applications at the user's request. This raises significant security issues, since many applications may contain malware. Even mobile platforms such as Android and iOS, which require all applications to be manually verified and vetoed by their respective app stores, are known to feature apps that contain malware. Google's Android ecosystem with its Play Store sets a particularly bad example with large amounts of malware applications featuring on the official store (Hong, 2015), however even Apple with its much more rigours app review system has fallen victim to allowing thousands of malware carrying apps (BBC, 2015) (McGarry, 2015).

For this reason, from a security perspective, the ideal option is to use dedicated hardware where only the LS software is allowed to run.

### 3.6.3. AIR-GAPPING WITH A DATA DIODE

It is a widely accepted fact that due to the complexities in both hardware and software no system can be considered 100% secure. However when designing the LifeServer one of the goals was to provide a truly secure data storage area that the users can rely on to keep their most private data. One way to achieve security in an unsecure system is to disconnect it from all networks, in what is called "air-gapping", a common strategy in systems such as those in nuclear power plants, military systems, and life-critical systems. There are known attacks on air-gapped systems (NSA, 2014), and these should be also addressed in our implementation.

An obvious problem with air-gapping the LS is that it would require a considerable amount of effort on the user to transfer data to and from the LS, and this would be unacceptable according to GGPs #1, #2, and #4.

In order to mitigate this problem, but still ensure that data on the LS cannot be compromised remotely over a network, the solution is to use a unidirectional network, also called a *data diode*.

A data diode is a physical link between two computers that only allows data to flow in one direction. Even if both computers try to reverse the flow of data, the data diode will prevent it from happening at the hardware level. Typical implementations of data diodes rely on custom fibre-optic cables, but such a solution would be prohibitively expensive for a consumer product. A solution will be provided for this problem.

### 3.6.4. DUAL-SYSTEM ARCHITECTURE

Having all the data stored behind a data diode, whilst ensuring its security, would prevent the Act and Mind APIs from functioning properly. Some of the data must be kept in a machine that has 2-way communication with the outside world. For this reason the decision was to design a dual-system architecture, with one computer connecting to the Internet and another air-gapped through a one-way data diode. The "public" Internet connected computer will run all the APIs and only send the most private of data to the "private" computer, where it will be stored for later use. The privately stored data can actually be of use in some conditions, which will be described later.

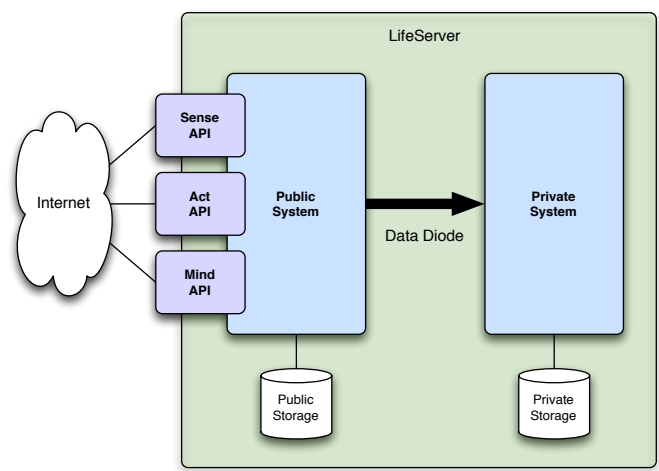

Figure 16 - A Dual-System Architecture

### 3.6.5. SEALED STORAGE AND PROCESSING

If use of a data diode ensures that no data can be transmitted out of the private RPi, which is a desirable security feature, the usefulness of that data will be very limited until it is moved from the private storage to another host where it can be processed and acted upon.

A compromise should be reached where the data can flow in both ways but in a controlled manner, dictated by a thin layer of software that polices what data travels from the private to the public RPi. Furthermore, if the private RPi creates a secret private key, and shares the corresponding public key to the user's devices, these devices can send highly private data to the private RPi, which can be saved on the private storage. As long as the user cannot access the private RPi's private key it can be said that the data is sealed beyond the user's reach. This is a very useful feature for those scenarios where the user wants to record and process data that would otherwise be illegal to so.

#### 3.6.5.1. SOLUTION FOR SOUND SENSING LEGAL ISSUES

For example, recording the user's surrounding sound 24/7 is an activity from which a great deal of context can be extracted (Cordeiro, 2013), however if the recording includes other people's conversations without their knowledge or consent it could be illegal under some jurisdictions. A legal definition of *sound recording* is *"a recording of sounds, from which the sounds may be reproduced"* (*Copyright, Designs and Patents Act*, 1988). Therefore it follows that as long as the user cannot reproduce the sounds, legally the activity is not considered to be sound recording and therefore is not illegal.

A solution to this problem would be for the recording device, such as a smartphone, to immediately encrypt the audio recordings with a private key which only the private RPi has access to, and sending those files to the private RPi. The user will not be able to playback the recordings even if he or she wanted to. At the same time, the private RPi will be able to process those files and extract useful context, which can be used by the Act and Mind APIs.

## 3.7. LIFESERVER SOFTWARE

This section details some of the decisions taken with regards to the LifeServer software. Similarly to the hardware section, the decisions will take into account the GGPs outlined in page 40.

### 3.7.1. DESIGN: API-FIRST VS UI-FIRST

Traditional user-centric systems normally start their development with requirements gathering followed by UI design. By presenting early mock-ups of the application interface to the users at the beginning of development cycle, this approach ensures early validation of the requirements.

Even though user-centric development is a proven methodology, with the arrival of IoT we are witnessing a greater focus towards machine-to-machine (M2M) systems that, although often serving users through end-point UIs, increasingly rely on well-designed APIs to interact with external systems. Providing well designed and open APIs also allows third-party developers to work on integrating their products and apps with the platform, which is akin to gaining development work at a low cost. For this reason tech giants such as IBM, Google and Apple are quickly adopting what is called *API-First Design* (Kapoor, 2015).

Grokya sits neatly as a middleware platform between consumers, developers and enterprises, and for that reason the logical choice is to follow the API-First Design method. The goal is to provide any developer with the freedom to develop great user-facing experiences *Powered-by-Grokya*, by drawing from the rich contextual insights made available by the system's APIs.

### 3.7.2. CLIENT-LS PAIRING

When sending personal data via the Sense API, the IoT device or app (herein referred to as clients) must ensure that it is sending the right data to the right LS. Data that is relevant to the user's interaction with said client is relevant and should be sent to the user's LS, and not to another user's LS by mistake.

This requires that Sense API clients established a logical link to the correct user, and this can be achieved explicitly or implicitly.

#### 3.7.2.1. EXPLICIT CLIENT-LS PAIRING

The easiest way to establish a pairing between a client and the LS is to have the user explicitly create that pairing, by selecting an option to that effect and providing a unique id that allows the client to find the LS on the network. A simple example would be to provide a unique username or domain name, which the client can use to lookup the LS on the network. Such a scheme would also require explicit authentication by the user, so that the client ensures the user does in fact have authority to create the pairing. OAuth 2 authentication can be used in such a scenario.

### 3.7.2.2. IMPLICIT CLIENT-LS PAIRING

The problem with the explicit strategy is that it goes against the GGPs in terms of intrusiveness. The *Grokya way* is to remain as invisible and unobtrusive to the user as possible.

Every sensor that captures data potentially relevant to the user's context should make that data available to the user's LS, in order to provide a more complete and accurate view of the user's context. Sensors include not only those controlled by the user but also those under the control of third parties, such as CCTV cameras, ATMs, geo-fencing beacons, RFID card readers, and every digital device that the user interacts with, directly or indirectly. An explicit pairing strategy would never work in such scenarios. It could be argued that another user's wearable sensors may provide context of relevance if the two users are in close proximity, but care should be taken that the data shared is in fact relevant and not in violation of a user's privacy. For example, if Alice and Bob were chatting to each other next to the coffee machine at work, it would be advantageous that Alice's Google Glass (or similar device) footage is shared with Bob's LS, providing a third-person view of him. The same cannot be said of Alice's heart rate, measured by her HRM monitor.

Sensors and LS must coordinate to determine which data is relevant to whom. It is equally important that private data is not accidentally shared with a user that is not in the vicinity of the sensor at that time, as that could potentially violate the privacy of other users.

### 3.7.3. MINIMUM ANDROID API VERSION

As part of this project's PoC, an Android app should be developed in order to test some of the APIs from the point-of-view of an IoT device.

When developing an Android application, a choice must be made with regards to the minimum Software Developer Kit (SDK), or Android API version, that the application requires. Choosing lower API versions means that more devices will support running the app, however it also means that some Android features will not be available to the app.

For the Grokya Android App the choice was API 18, which corresponds to Android 4.3 (Jelly Bean), because it is the lowest API that supports Bluetooth Low Energy (BLE) connectivity. BLE is essential for connecting to external wearable sensors, and beacons, whilst keeping battery consumption to a minimum. By choosing API 18, at the time of writing (July 2015) an estimated 40.9% of the devices that are active on the Google Play Store will be able to run this app. This number is expected to increase as more device owners upgrade their devices.

### 3.7.4. RANDOM NUMBER GENERATOR: PSEUDO VS TRUE

A secure system requires strong cryptographic keys that in turn require an equally strong random number generator. Random number generators are typically divided into two types: Pseudo-Random Number Generators (PRNGs), and True Random Number Generators (TRNGs).

PRNGs use a mixture of mathematical formulas and pre-defined tables to generate large amounts of seemingly random numbers. However the process is deterministic and, even if very efficient, the numbers are not truly random.

TRNGs, on the other hand, make use of unpredictable external physical phenomena in order to generate true random numbers. Since the process depends on the limited amount of entropy generated by its physical input, TRNGs are considerably slower than TRNGs.

Even though TRNGs are less efficient, for the purposes of this project, the need is for strong cryptographic keys, therefore quality is preferred over quantity. For this reason the system should make use of a TRNG, and section 5.2 will include details of two TRNGs tested for the project.

## 4. IMPLEMENTATION

Due to the relatively big scope of this project, a full implementation of every aspect of the platform would be impossible given the limited amount of time available. Choices had to be made with regards to the scope of implementation.

### 4.1. IMPLEMENTATION SCOPE

The main focus of the implementation was on building a secure hardware platform on which the LS software and APIs could execute.

The secondary focus was on the actual APIs and providing a POC of how VDP can be applied to a Mind API operation. Although originally considered in scope, the API implementation was not completed.

Other aspects such as a peer-to-peer network of LS nodes and Mind API operations over large number of nodes were left out of scope.

### 4.2. DEVELOPMENT LIFECYCLE

This project has followed a prototyping software development lifecycle, which is summarised in Figure 17 below.

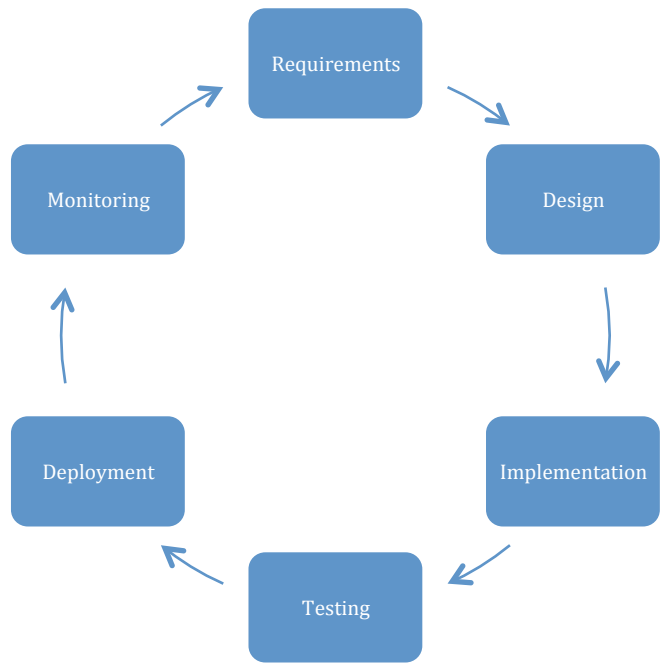

Figure 17 – Prototyping Development Lifecycle

Since the overall POC relied on multiple independent systems working together, the interoperability between these systems could present significant unknown obstacles and therefore posed the highest implementation risk. For this reason the early focus of the implementation was to concurrently develop the basics of each system to a point where their interaction could be proven to work. Only then was each system independently developed to fulfil its individual requirements and design specifications.

The implementation started with the hardware, followed by the software. The following sections describe details of each.

## 4.3. HARDWARE IMPLEMENTATION

For the purpose of the project's POC the choice of hardware was relatively straightforward. The Raspberry Pi (RPi) computer is both affordable and great for prototyping. It features a General Purpose I/O (GPIO) connector that will also come in useful to connect the two systems via the data diode.

A 1TB external USB hard disk drive (HDD) was chosen to provide the private RPi with extra storage, since this is where most of the long-term storage will reside.

A smaller 128GB USB flash drive was chosen to provide the public RPi with extra storage.

The main components for the hardware prototype are as follows:

- Public RPi: Raspberry Pi 2 Model B

- Private RPi: Raspberry Pi Model B+

- Public RPi storage: SanDisk 128GB Flash Drive

- Private RPi storage: Western Digital Elements 1TB USB HDD

- USB Hub: Transcend TS-HUB3K USB 3.0 Hub

The following diagram details the hardware components and how they link to each other.

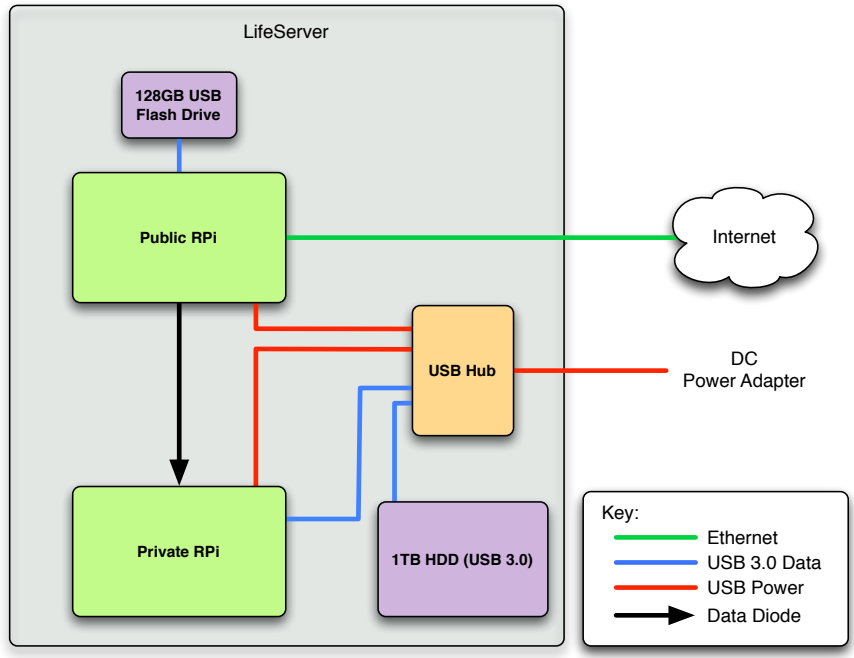

Figure 18 - LifeServer Hardware Layout

A powered USB hub was included to provide power for both the RPis as well as for the external HDD, via a USB 3.0 cable.

The reason why the Private RPi cannot connect directly to the external HDD is because the later draws more current than the Private RPi can provide directly via it's own USB port.

On the other hand, the Public RPi's external storage can be connected directly, because the Flash Drive has much lower power requirements than an external HDD.

The public PI is the only component that connects to the external network.

Even though both the public and private RPis connect to the same USB hub, they public RPi only does so via it's power-in USB jack. This means that no data communication can take place between the 2 RPis via the USB hub. Theoretically, if the RPi board could measure variations in it's input power voltage, some communication could take place, but that is not the case with the current RPi hardware.

### 4.3.1. DATA DIODE

The data diode component deserves especial attention, since it's a key element of the hardware. Its purpose is to ensure that no data can travel from the private RPi to the public RPi, even in the case that both RPis are colluding to make such a data transfer. The private to public data transfer must be impossible at a physics level on the wire.

#### 4.3.1.1. **PROTOTYPE 1**

Following the prototyping methodology the first implementation simply made use of a wire connecting the serial transfer (TX) from the public RPi directly to the serial receive (RX) of the private RPi. The only other connection was the ground (GND) between the two RPis. By missing out the reverse connection (private TX to public RX) it prevented data from travelling in the reverse direction. Early tests using the command line to send data through the serial connection showed that it worked as intended.

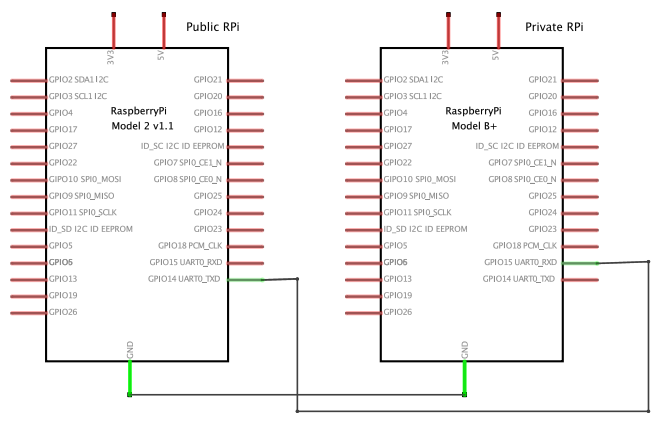

Figure 19 - Prototype 1 of the data diode

4.3.1.2. **PROTOTYPE 2**

However theoretically it was still possible for the RPIs to programmatically reconfigure and reverse the TX/RX functions of the GPIO pins, defeating the purpose of the data diode. The solution was to add a physical IC between the 2 RPis in such as way as to physically prevent the reversal of the data flow on the wire. For this purpose a non-inverting buffer was chosen, the CD4050. Prototype 2 was wired in the following way.

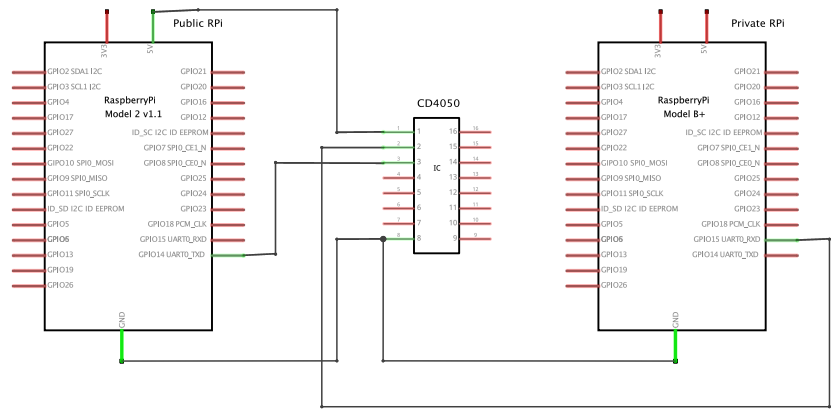

Figure 20 - Prototype 2 of the data diode

This setup was tested; including swapping around the TX/RX on the RPi units (TX on the private RPi and RX on the Public RPi) to ensure the IC buffer was doing its job. As expected, no communication was possible from the Private RPi to the public RPi. The hardware data diode implementation was finalised.

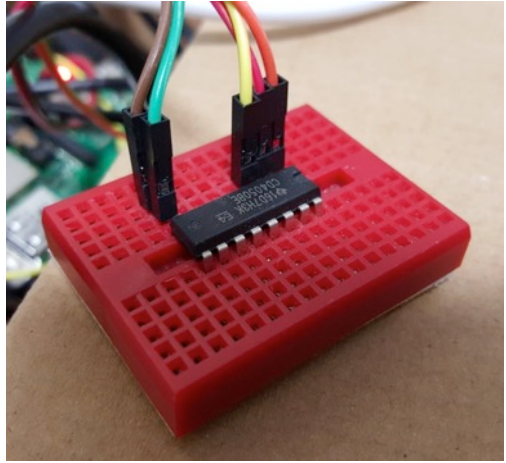

Figure 21 – Detail of final data diode implementation

### 4.3.1.3. FORWARD ERROR CORRECTION (FEC)

It should be noted that unidirectional communications prevent the receiver from sending simple acknowledgements of data received back to the sender. If data corruption occurs the sender will have no way of knowing that it should resend the data. In order to work around this issue, a technique called forward error correction (FEC) can be used, where the sender adds redundancy to the data, allowing the receiver to reconstruct the original data even if noise is introduced in the data during the transfer. This redundancy in encoded data is normally found in common household items such as CDs, DVDs and QR codes, which allow the readers to deal with varying degrees of damaged media, like scratches in a DVD or partially covered QR codes. The technique is also applied in long-range communications such as deep-space probes that transmit data back to Earth. Reed–Solomon codes, turbo codes, and low-density parity check codes are three of the most common FEC methods.

After extensive tests, even at baud rates of 115200, no errors were detected in data transmission between the two RPis. For this reason it was decided that FEC was not necessary for the purpose of the POC. However if at a later stage the speed of transmission is to be increased (with Linux kernel changes it can reach mega baud rates) then FEC should be implemented.

**4.3.2. ELECTROMAGNETIC ISOLATION (FARADAY CAGE)**

As mentioned in the literature review, a typical side-channel vector of attack is the gathering of computer internal state by analysing minute electromagnetic variations emanating from said computer. In the case of the RPi this issue is magnified because it ships with an FM transmitter out of the box, making it trivial to broadcast data wirelessly to within a few centimetres or, provided it is fitted with a wire on GPIO pin 7 acting as an antenna, to within a few meters. Since we cannot trust the private RPi's underlying stack, we need to completely isolate it from an EM point of view, and the solution here is to enclose it in a faraday cage. When choosing the material for the faraday cage the most important factor is its conductivity, with the highest conductivity being the best. Table 2 shows the relative conductivity of different types of material.

| Material IACS | % Conductivity |
|---|---|
| **Silver** | 105 |
| **Copper** | 100 |
| **Gold** | 70 |
| **Aluminum** | 61 |
| **Nickel** | 22 |
| **Zinc** | 27 |
| **Brass** | 28 |
| **Iron** | 17 |
| **Tin** | 15 |
| **Phosphor Bronze** | 15 |
| **Lead** | 7 |
| **Nickel Aluminum Bronze** | 7 |
| **Steel** | 3 to 15 |

Source: ("Electrical Conductivity of Materials - Blue Sea Systems," n.d.)
Table 2 - IACS conductivity of common materials

Given this information, the choice was made to use copper fabric.

A template was made using the online tool Template Maker[1] with the dimensions of the enclosure where the private RPi was going to be inserted. After printing and transferring the template to the copper fabric, it was cut and folded along the correct places. Finally it was placed around the RPi, ready for the next step in the implementation.

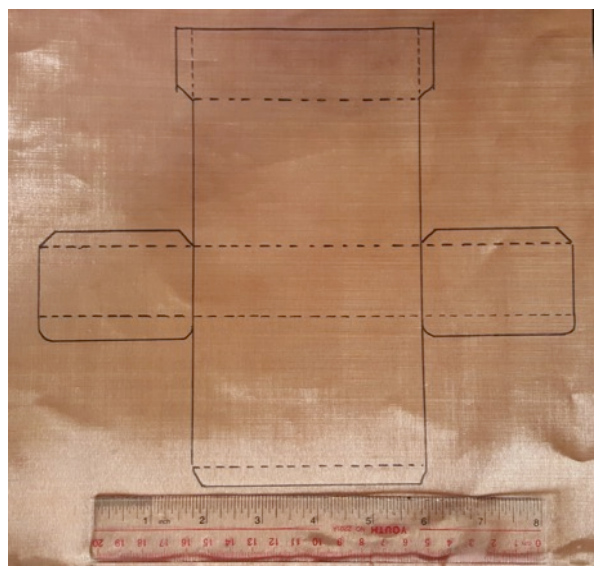

Figure 22 - Faraday cage template

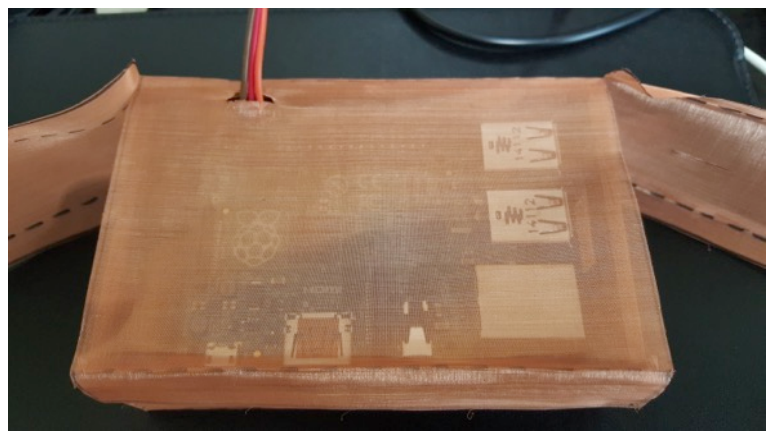

Figure 23 - Faraday cage installation around RPi

---

[1] http://www.templatemaker.nl

### 4.3.3. THERMAL ISOLATION

Another side-channel attack mentioned in section 2.3.5.1, page 34, was that of data leakage through thermal variations on one machine being sensed by another nearby machine. Tests were performed to analyze the feasibility of such an attack between 2 RPis located in a small enclosure. The full test results are located and discussed in section 5.1. In summary, the feasibility of such an attack on the proposed setup is very low and no specific measures need be taken for this purpose alone. However, as will be discussed later, the potting of the private RPi for different reasons actually helps to protect against these kinds of thermal attacks.

### 4.3.4. SEALING THE PRIVATE RPI (POTTING)

In order to allow for the sound sensing use case described in section 3.6.5.1, the solution was to totally seal the private RPi in a way that makes it nearly impossible for the user to access it's components, specifically the micro-SD card where the RPi stores the private key against which all the sensed data is encrypted. This way, even if the user has access to the private storage HDD used by the private RPi, without the private key to decrypt the data he or she cannot reproduce the sound files, therefore avoiding the legal definition of sound recording.

The most cost effective method of sealing the private RPi is to perform what is called *potting*. Potting an electronics assembly is the process of totally filling the assembly with a material, such as silicone or epoxy. When the material dries it forms a strong enclosure that protects the assembly from water, humidity, and dust, but more importantly, from direct access by a person to the assembly itself.

In order to perform the potting of the private RPi, a special container had to be designed in order to allow a generous amount of material to surround the RPi, especially the zones to the sides and under the RPi where the micro-SD card is located. Small holes had to be included to allow the running of the power cable and usb cable for the external HDD. The data diode cable did not require a hole in the case because it runs vertically, out of the GPIO.

The case was designed in SketchUp 3D[1] and two prototypes were made. The first prototype was not suitable because the holes were not big enough to allow for the two USB connectors to reach the RPi, and the four support columns for the RPi were too narrow and fragile. The second prototype was revised, in increased holes and larger supports. The final 3D model can be seen below.

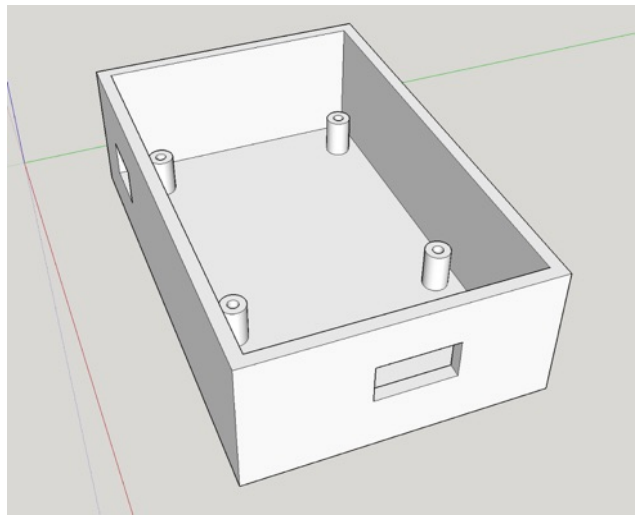

Figure 24 - 3D Model of RPi Container for Potting

This 3D model was then exported as an STL file and 3D printed with PLA.

---

[1] http://www.sketchup.com/

The final enclosure, with the RPi already mounted and the faraday cage fabric semi-installed can be seen in the following image.

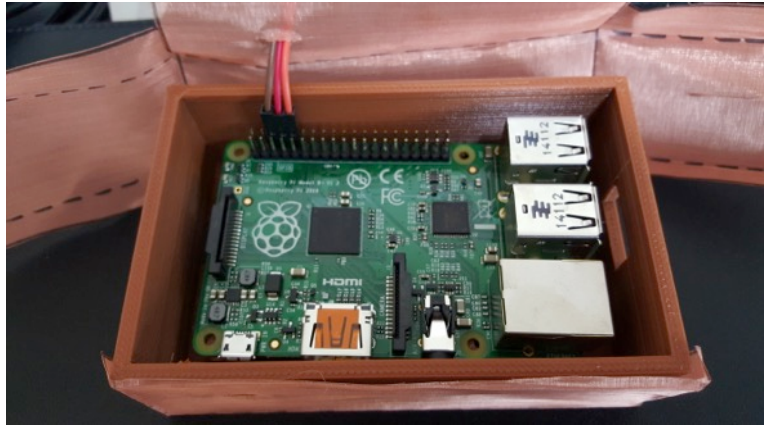

Figure 25 - RPi in 3D printed container

Before mounting the RPi to the enclosure the micro-SD card, already pre-loaded with the basic OS image, was inserted into the RPi and super-glue was applied to it, making it virtually impossible to remove the micro-SD card from the RPI without risking damaging it, which is the point of the exercise.

With the power cable, USB cable for external HDD, and the data diode cable all attached, the assembly was ready for the potting.

Care was taken to apply hot-glue to the gap between the holes and the cables, however this was done on top of the faraday cage fabric, which proved to be a mistake. The hot glue did not seal the holes properly and the assembly was not water tight, which is obviously an important requirement when potting.

For this potting the following epoxy compound was chosen: YH-9002A/B.

The epoxy comes in two containers, one with the resin and another with the hardener, and must be mixed in a 5 to 1 ratio by weight, exactly. Once mixed it can be applied to the assembly, first with the use of a large syringe for pushing the epoxy under the assembly, and later just by pouring it directly from the mixture container.

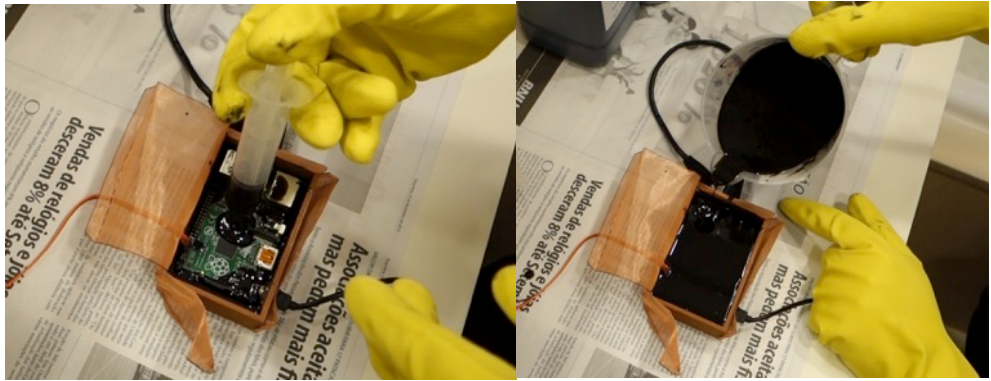

Figure 26 - Potting RPi with epoxy

Once finished the epoxy will take up to a few days to cure and solidify, depending on the temperature. Once cured, it is extremely hard to remove.

The final appearance of the private RPi, after potted and sealed with the faraday cage:

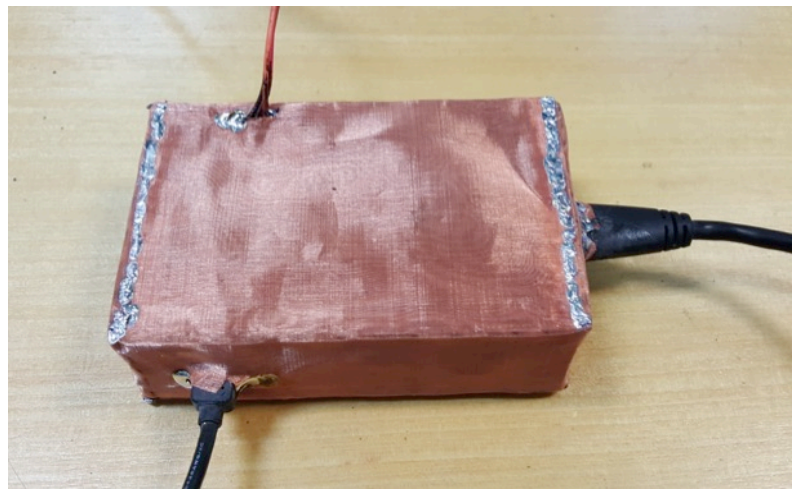

Figure 27 - Private RPi potted and with faraday cage

### 4.3.5. FINAL HARDWARE ASSEMBLY

With the private RPi potted and totally sealed with the faraday cage, the final assembly took place. A shoebox was chosen as the enclosure for the system. Also a passive PoE cable was used in order to have a single port on the enclosure that provides both power and network to the whole system.

Due to the limited space in the enclosure, the RPi's external HDD was placed at the bottom of the enclosure, with the 2 RPi's placed on top of it.

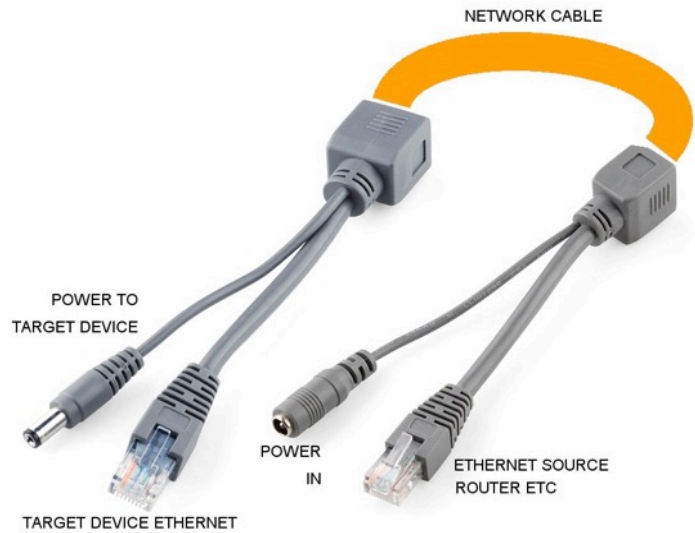

Figure 28 - Passive PoE set

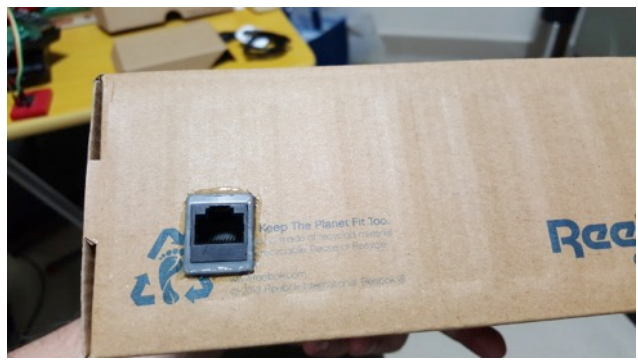

Figure 29 - Passive PoE port open on enclosure

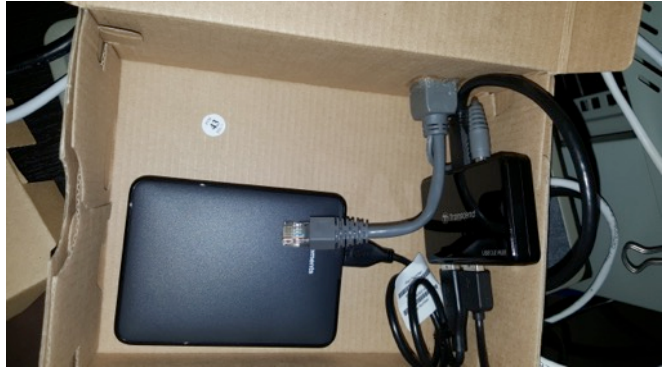

Figure 30 - Initial installation of HDD and USB hub

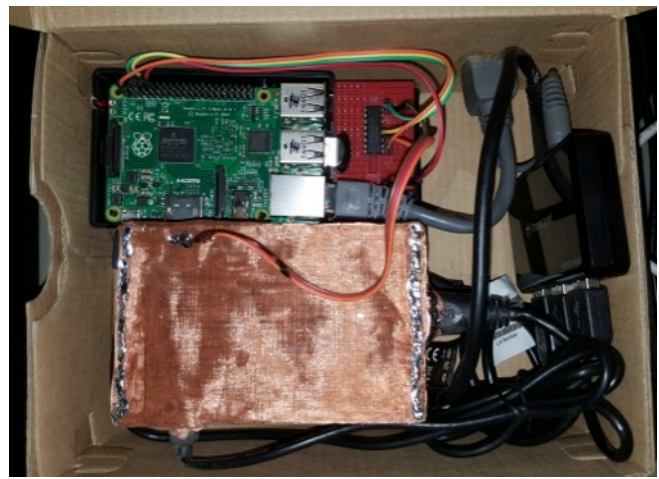

Figure 31 - RPis and data diode installed and running

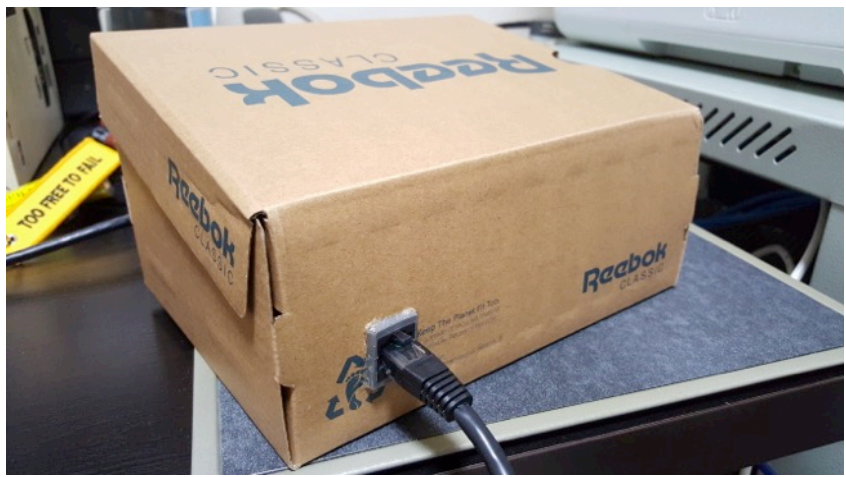

Figure 32 - The LS up and running

## 4.4. SOFTWARE IMPLEMENTATION

With regards to the software the implementation was split into three parts:

1. Setting up the base operating system and required dependencies
2. Develop a service for Public RPi to Private RPi intercommunication, called Callosum
3. Develop a basic implementation of the APIs, on the Grokya Server App.

The architecture of the software stack is as follows:

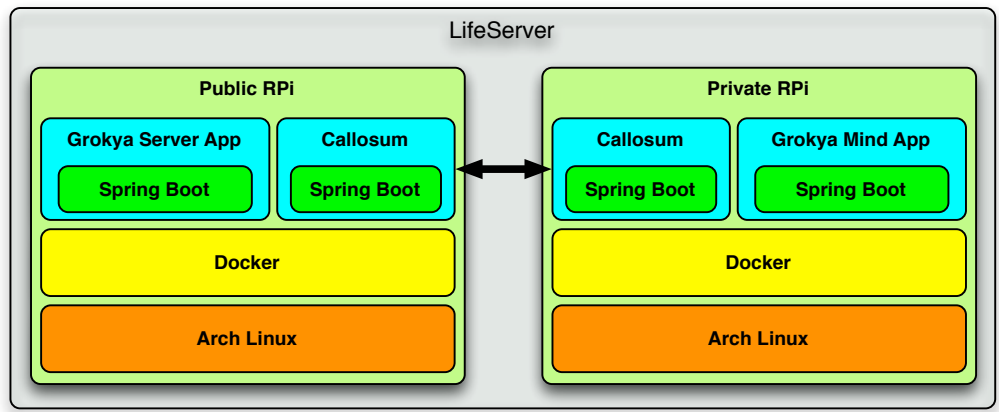

Figure 33 - LS Software Stack

From bottom to top, the OS is Arch Linux, on top of which Docker runs multiple software containers, the actual LS services.

Docker was chosen as opposed to heavier virtualisation technologies that would be unfeasible on a RPI. This approach is called a micro-service architecture, which favours multiple small single-purpose apps rather than one big monolithic application. The benefits are ease of maintenance and the extra security inherent to the compartmentalisation of the services.

The main software apps are:

- Grokya Server App: runs on the public RPi and is responsible for the 3 main public facing APIs.
- Callosum: runs on both RPis and is responsible for controlling the communication between the two RPIs. It should be noted that this service deals both with a data-diode scenario as well as with a 2-way communication channel and can interchange between both modes during runtime.
- Grokya Mind App: (not to be confused with the Mind API) should on the private RPi and be responsible for the sealed processing of the data it stores. It should be able to respond to queries from the public RPi in the 2-way communication scenario. Due to lack of time, at the time of writing, this app was not yet developed.

### 4.4.1. OPERATING SYSTEM

Arch Linux was chosen as the operating system because it is lightweight and relatively secure, whilst also featuring a wide choice of packages easily installable via its package management tool *pacman*.

The main focus in terms of installation was to develop a repeatable procedure that, given the same starting point with a base image of Arch Linux, would always arrive at a fully configured LS system. Instead of just documenting the procedure, shell scripts were created to automate the process. This way the shell scripts also act as documentation, and provide transparency to the whole process, which would not be the case if a fully pre-configured binary image were made available instead. The shell scripts can be found in Appendix 10, page 98.

### 4.4.2. GROKYA SERVER APP

The Grokya Server app partially implemented the main APIs as well as a basic VDP payment processor that distributes revenue amongst the stakeholders (i.e. Sense API data sources) that contributed to that revenue. As with all the other apps, it was written in Java 8 and used Spring Boot as the underlying framework because it is very lightweight and suited to a micro-service running within a docker container.

### 4.4.3. CALLOSUM

The Callosum app implemented a simple wire protocol for the intercommunication between the two RPIs. It introduced the concept of a CallosumPacket that can encapsulate several types of messages, from a pre-defined set of message types. When sending data, the sending RPi serializes the CallosumPacket and sends it via the wire to the other RPi, which receives it and deserializes it. Depending on the message type, the receiving RPi will route de message to the appropriate service.

### 4.4.4. SOFTWARE DEVELOPMENT PROCESS

The software was developed using the Eclipse IDE, and is version control with the use of Git. When developing features, the app is first tested within Eclipse, as well as a docker container in the developer's machine. When it runs satisfactorily, a git push sends the changes to a central repo.

Since docker images are architecture specific, a docker image created in a Mac cannot run on the Raspberry Pi. For this reason a build system was setup with a RPi dedicated to build the software for the correct architecture.

## 5. TESTING

Several tests were performed during the prototyping process. This chapter details the tests and their results.

### 5.1. THERMAL ATTACK TESTS

In order to test the feasibility of a thermal attack as the one described in section 2.3.5.1, a controlled environment was setup, using temperature sensors running on Arduino Uno boards. The setup was as follows:

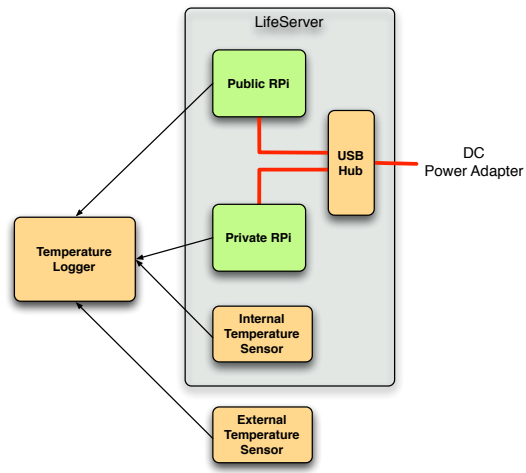

Figure 34 - Thermal Test Setup

The Arduinos and the RPis all ran a small web-service that, upon requested by the logger, returned the current temperature reading. In the case of the Arduinos, they returned the reading from the DHT22 temperature and humidity sensor that they were fitted with, whereas the RPis returned the reading from their internal CPU temperature sensor.

During the test, the public RPi was idle, and the private RPi was regularly throttling its CPU on and off, by running the command "yes > /dev/null" concurrently 4 times, in order to max-out all 4 cores in the RPi 2 (these tests were conducted with 2 RPis 2). The hypothesis is that by manipulating the CPU load on one of the RPis, the resulting increase and decrease in heat will be detected by the other RPi's build-in sensor. Several tests were performed, each test using a different frequency in the CPU throttling cycles.

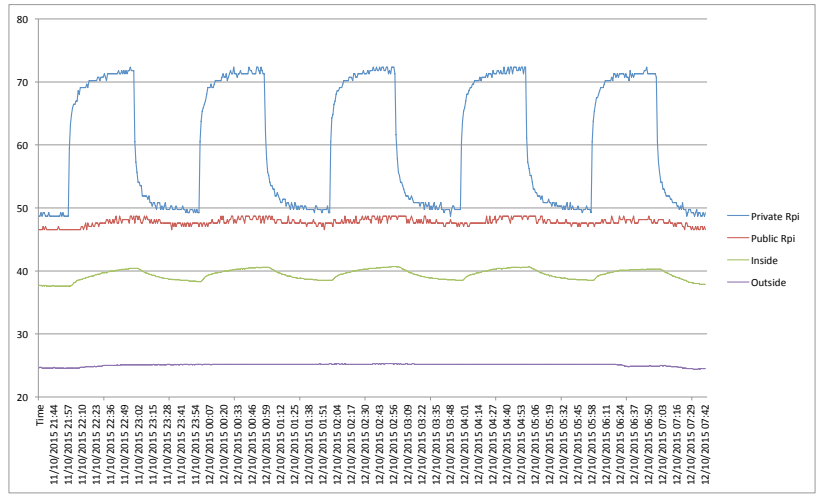

Figure 35 - CPU throttling generated heat: 1-hour cycles

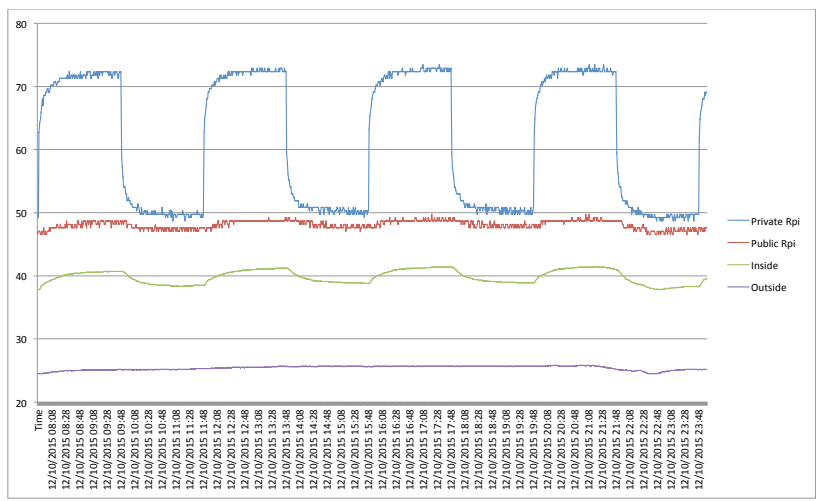

Figure 36 - CPU throttling generated heat: 2-hour cycles

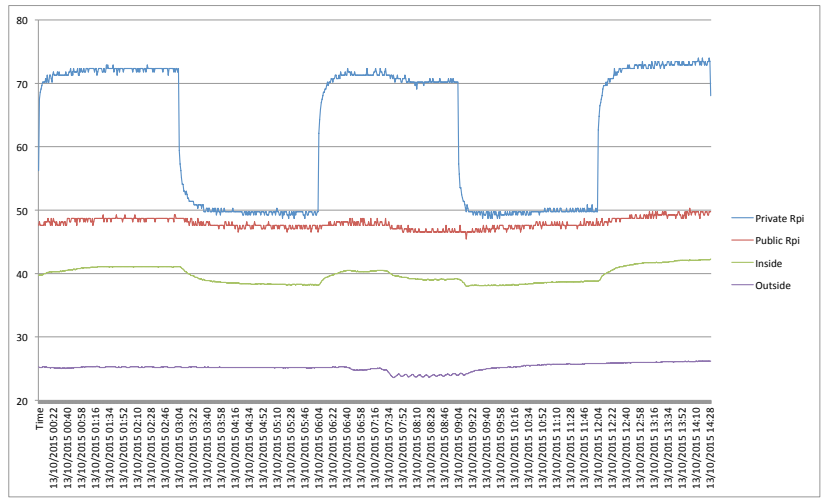

Figure 37 - CPU throttling generated heat: 3-hour cycles

The tests show that, despite a significant oscillation in the temperature of the private RPi, the public RPi hardly moves, with an average of 1 degree Celsius deviation. Taking into account that any processing activity on public RPi or outside temperature changes would have a significant impact on the public RPi temperature, it is safe to conclude that a thermal attack on this current setup is not viable. Furthermore, the final LS setup uses a Raspberry Pi B+ as private RPi, meaning that it has only 1 CPU and therefore less thermal impact.

Finally, the potted RPi runs considerably cooler than a non-potted RPi, probably because the potting material acts as a heatsink.

## 5.2. RANDOM NUMBER GENERATION TESTS

Given the need for a good source of entropy and randomness, two TRNGs were tested and compared:

- the RPi's built-in hardware random number generator (present in both the BCM2835 SoC of the models A, B, B+ and Compute Module and the BCM2836 SoC of the Raspberry Pi 2 Model B);

- the NeuG TRNG running on an FST-01 USB dongle computer, a fully free (as in freedom) platform.

Both TRNGs were tested using the well-known random number testing suite Dieharder (Brown, Eddelbuettel, & Bauer, 2009). The full results of the tests are presented and discussed below.

**Setting up Raspberry Pi built-in TRNG**

The built-in hardware device on the RPi does not require any special setup on Arch Linux, and it should be noted that it is already configured by default to supply entropy to `/dev/random`.

**Setting up NeuG device**

The NeuG hardware device does not require any special drivers for Linux. Simply plugging it into one of the USB ports will create a TTY device identifier, in this case `/dev/ttyACM0,` that outputs a stream of random data on demand.

It should be noted that the TTY device does need some configuration, to act as a raw device. Failing to do so would result in premature end of streaming due to programs interpreting random "end-of-file" (EOF) signals. The correct TTY configuration, as per the NeuG documentation, should be set as follows:

```
$ stty -F /dev/ttyACM0 -echo raw -parenb
```

All random number files were truncated to the same size, so that the dieharder test suite rewinds them the exact same number of times during the tests.

```
dd if=dev-urandom.raw of=dev-urandom-trunc.raw bs=1M count=7168
```

The commands to run all the tests was:

```
time dieharder -g 201 -a -f dev-urandom.raw > dieharder-test-dev-urandom.txt
time dieharder -g 201 -a -f dev-random.raw > dieharder-test-dev-random.txt
time dieharder -g 201 -a -f neug-random.raw > dieharder-test-neug-random.txt
time dieharder -g 201 -a -f pi-builtin-random.raw > dieharder-test-pi-builtin-random.txt
time dieharder -g 201 -a -f pi-builtin-random-big.raw > dieharder-test-pi-builtin-random-big.txt
```

The results, which can be found in Appendices 11 through to 14, reveal that no generator stands out from the others. The main difference is that NeuG is limited to around 40Kbps output, whereas the RPI's built-in works must faster, at rates that exceed 100Kbps. For that reason the decision is to use `/dev/random` which draws entropy from the built-in generator.

# 6. CONCLUSIONS

This chapter will firstly report some of the problems found during the project and how they were resolved.

It will then discuss the implications of the work and provide recommendations for future work.

## 6.1. PROBLEMS ENCOUNTERED

### 6.1.1. SLOW BUILD TIMES ON RASPBERRY PI

It should be said that the build process on the RPi is very time consuming. A build that takes a mere 12 seconds on a MacBook Pro, takes in excess of 20 minutes on a Raspberry Pi 2. Fortunately the subsequent running of the software doesn't experience such a similar latency. At present there is no solution to this problem, except for using a more powerful build machine and cross-compiling to the arm architecture.

### 6.1.2. POTTING A NON-WATER TIGHT CONTAINER

In the preparation for the potting of the private RPi, a mistake was made when making the container watertight. The two small openings for the power and USB cable were filled with hot-glue, however this procedure was done on top of the faraday cage fabric that was already in place. This means that the hot glue did not make full contact with the inside of the holes, leaving small gaps. When filling the container with the epoxy these holes, albeit small, leaked a significant amount of epoxy leading to a fairly messy finish on the outside. The faraday cage was actually ruined and was replaced later. The leaked epoxy eventually cured and was removed with the aid of a power tool. It should be said that it was fairly hard to cut the excess epoxy, which is a good sign in terms of the security afforded by the potting.

## 6.2. DISCUSSION

### 6.2.1. CONTEXT-AWARE SECURITY

Context-aware systems can offer a great alternative to traditional authentication systems that typically rely solely on user memorized passwords. Users often chose passwords that are trivially easy to guess resulting in poor security. Similarly to the Nymi biometrics bracelet mentioned in page 22, a context-aware system that keeps track of the user's context in real time and 24/7 can be queried by external systems regarding the probability that the user is indeed likely to be attempting to authenticate against the system in question. For example, using an ATM can be made safer if the ATM queries the user's LS to verify that the user is in fact, using the ATM (or the likelihood of it). If the user is in a different location or context (i.e. sleeping) then the ATM operation should be flagged as highly suspicious.

## 6.3. MOVING AWAY FROM ADVERTISING

Advertising-based revenue models became popular with centralised web services because the Internet lacked a built-in payment system. This project proposed a decentralised methods of distributing value, so it follows that the online advertising model should now be challenged. If a decentralised micro-payment network makes ads redundant, then the usage of Grokya will also shift away from advertising and towards other kinds of marketing activities.

## 6.4. THE INTERNET OF MINDS

If the goal of an LS is to become the "digital mind" of its user then it should follow that a p2p network of LS devices becomes an Internet of Minds that interacts with the Internet of Things. This can have a significant impact in many aspects of life, not least governance.

## 6.5. THE ISSUE OF TRUST

As long as an IoT device is under the control of a single entity (i.e. the manufacturer), with full control over its software updates, it is perfectly possible for that device to be turned into a tool of surveillance at any moment. The only way to avoid such a situation of power over the user is to relinquish that position of power by ensuring that the IoT device only runs open source code co-signed by a decentralised authority, in the shape of a committee of industry experts who can review and approve the software updates. Only by making the whole process of software updates transparent can the manufacturers give the users the assurance that their device is acting on their behalf and not against them.

## 6.6. TOWARDS DIRECT DEMOCRACY

It would not be far fetched to imagine a world where citizens do not elect representatives every few years, but instead have a direct vote in each and every decision that affects their lives. Of course voting several times a day would require a significant effort and one that most people probably would not be able to commit to. However if an LS reaches a high level of understanding of its user and their needs and wants, then hypothetically the LS could vote on behalf of the user in some, if not all, the micro-referendums.

## 6.7. THE FUTURE OF LAW ENFORCEMENT

Albeit it being a potentially controversial topic, the fact is that governments and law enforcement agencies do not share the same economic incentives as for-profit enterprises to track citizens. It could be argued that, given some definition of the law in terms of code, if an LS knows everything about its user then it can be in a position to judge whether the user is staying on the right or wrong side of the law. This can have huge implications, because just as enterprises do not need to track consumers in order to benefit from high-level analytics, then neither do the law enforcement agencies need to closely track our every movement in order to weed out the bad form the good. On the other to have a user-owned device potentially reporting its owner to the authorities, regardless of the crime, raises moral issues. Who decides on the rules, and who decides on their translation into machine-readable code?

This thesis will conclude with a quote that, in one sentence, both elegantly and bluntly summarises the vision and motivation behind this project:

*"If someone is going to spy on you,*

*it's probably best if its you."* – Fred Wilson (2006)

# 8. APPENDIX – RASPBERRY PI B+ MECHANICAL SPEC

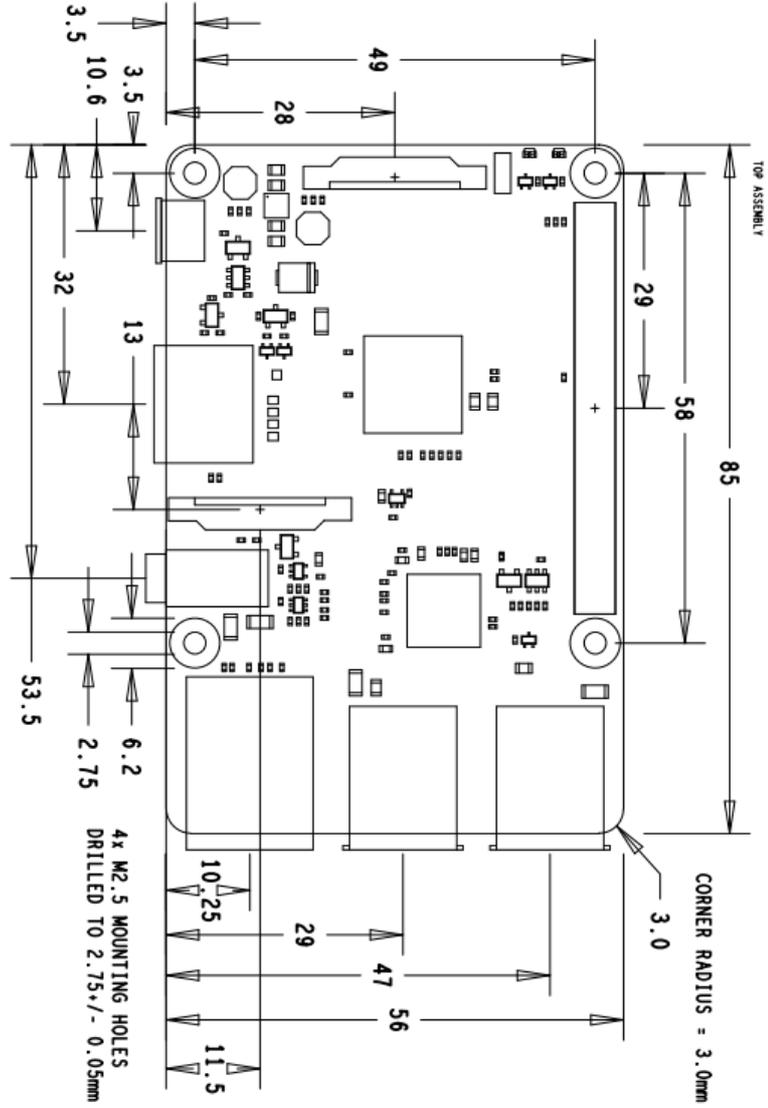

## 9. APPENDIX – POTTING COMPOUND SAFETY DATA SHEET

物 质 安 全 资 料 表

MATERIAL SAFETY DATA SHEET

一、化学品及企业标示 Chemical product and company identification

| |
|---|
| 物品名称：环氧树脂胶水<br>Article： |
| 规格型号：YH-9002A/B<br>Spec： |
| 制造商或供货商名称：广州市铂桥电子材料有限公司<br>Manufacture or supplier: |
| 地址及电话：广州市南沙区东涌镇大同工业区<br> Address:<br>TEL: 020-34903825  FAX：020-34903825 |

二、成分资料 Composition and information on ingredients

| 物品中（英）文名称：EPOXY   TH100A/B | | | |
|---|---|---|---|
| 同义名称：环氧树脂  $C_2H_3O$-$[C_{18}H_{20}O_3]C_{15}H_{14}O_2$-$C_2H_3$ | | | |
| 中文名称 | 化学式 | CAS.NO | 比例 |
| 1、环氧树脂 | $C_2H_3O$-$[C_{18}H_{20}O_3]C_{15}H_{14}O_2$-$C_2H_3$ | 25068-38-6 | 52% |
| 2、环氧稀释剂 | R-$C_2H_3O$ | 68609-97-2 | 14% |
| 3、固化剂 | $CH3(CH_2)$-CH=CH-$CH_2$-CH($CH_2$) | 693-98-1 | 25% |
| 4、添加剂 | $H_2NC_3H_6Si(OC_2H_5)_3$ | 108-88-3 | 8% |
| 5、色料 | C | 1333-86-4 | 1% |

# 10. APPENDIX – LS OS PREPARATION SHELL SCRIPTS

**01-grokya-ls-sd-preparation.sh**

```bash
#!/bin/bash

# set abort on error
set -e

hdd="$1"

cd ${0%/*}/..

echo "Partitioning SD card..."

echo "o
n
p
1

+100M
t
c
n
p
2

w
"|fdisk $hdd

echo "Formatting boot partition..."

yes | mkfs.vfat ${hdd}1
mkdir -p work/mount/boot
mount ${hdd}1 work/mount/boot

retval=$?
if [ $retval -ne 0 ]; then
    echo "Return code was not zero but $retval"
fi

echo "Formatting root partition..."

yes | mkfs.ext4 ${hdd}2
mkdir -p work/mount/root
mount ${hdd}2 work/mount/root

echo "Installing Arch Linux..."

mkdir -p work/downloads

echo "Downloading Arch Linux if necessary..."
wget -nc -P work/downloads/ http://archlinuxarm.org/os/ArchLinuxARM-rpi-2-latest.tar.gz

echo "Extracting Arch Linux..."
bsdtar -xpf work/downloads/ArchLinuxARM-rpi-2-latest.tar.gz -C work/mount/root
sync
echo "Moving /boot files in place..."
mv work/mount/root/boot/* work/mount/boot

echo "Installing startup scripts..."
rsync -av ../ls-setup-scripts work/mount/root/root/ --exclude work/mount

echo "Starting chroot phase..."
./scripts/02-grokya-ls-chroot.sh

echo "Umounting working directories..."
umount work/mount/boot work/mount/root
```

```
echo "Done!"
```

### 02-grokya-ls-chroot.sh

```
arch-chroot work/mount/root/ su - root -c /root/ls-setup-scripts/scripts/03-grokya-ls-arch-setup.sh
```

### 03-grokya-ls-arch.sh

```
#!/bin/bash

# set hostname
echo lifeserver > /etc/hostname
sed -i 's/localhost/lifeserver/g' /etc/hosts

# add user
useradd -m grokya
echo "grokya:grokya" | chpasswd

# update system
pacman --noconfirm -Syu

# install java and rxtx
pacman --noconfirm -S jre8-openjdk java-rxtx

# setup serial port
usermod -aG lock,uucp grokya
# TODO: the next 2 lines may have to be run on first boot (chroot doesn't seem to set them properly)
stty -F /dev/ttyAMA0 -echo raw -parenb 115200
ln -s /dev/ttyAMA0 /dev/ttyS80

# install docker
pacman --noconfirm -S docker
usermod -aG docker grokya
systemctl enable docker

# fix docker for RPi (https://github.com/docker/docker/issues/16256)
sed -e "s/ExecStart=\/usr\/bin\/docker -d -H fd:\/\//ExecStart=\/usr\/bin\/docker -d -H fd:\/\/ --exec-opt native.cgroupdriver=cgroupfs/" /usr/lib/systemd/system/docker.service > /tmp/docker.service
mv /tmp/docker.service /usr/lib/systemd/system/docker.service

# fix IPv6->IPv4 binding issue (https://github.com/docker/docker/issues/2174#issuecomment-149803815)
echo "IPMasquerade=yes" >> /etc/systemd/network/eth0.network
```

# 11. APPENDIX – DIEHARDER TEST: /DEV/URANDOM

```
$ dieharder-test-dev-urandom.txt
#=============================================================================#
#            dieharder version 3.31.1 Copyright 2003 Robert G. Brown          #
#=============================================================================#
   rng_name    |           filename             |rands/second|
 file_input_raw|              dev-urandom.raw|  1.02e+07  |
#=============================================================================#
        test_name   |ntup| tsamples |psamples|  p-value |Assessment
#=============================================================================#
   diehard_birthdays|   0|       100|     100|0.01601781|  PASSED
      diehard_operm5|   0|   1000000|     100|0.81220319|  PASSED
  diehard_rank_32x32|   0|     40000|     100|0.27361261|  PASSED
    diehard_rank_6x8|   0|    100000|     100|0.05141190|  PASSED
   diehard_bitstream|   0|   2097152|     100|0.27894532|  PASSED
        diehard_opso|   0|   2097152|     100|0.60673609|  PASSED
        diehard_oqso|   0|   2097152|     100|0.25737640|  PASSED
         diehard_dna|   0|   2097152|     100|0.20130100|  PASSED
diehard_count_1s_str|   0|    256000|     100|0.92825850|  PASSED
diehard_count_1s_byt|   0|    256000|     100|0.28826158|  PASSED
 diehard_parking_lot|   0|     12000|     100|0.86481685|  PASSED
    diehard_2dsphere|   2|      8000|     100|0.56912624|  PASSED
    diehard_3dsphere|   3|      4000|     100|0.30329621|  PASSED
     diehard_squeeze|   0|    100000|     100|0.99380636|  PASSED
        diehard_sums|   0|       100|     100|0.49104912|  PASSED
        diehard_runs|   0|    100000|     100|0.11960026|  PASSED
        diehard_runs|   0|    100000|     100|0.24135654|  PASSED
       diehard_craps|   0|    200000|     100|0.87326261|  PASSED
       diehard_craps|   0|    200000|     100|0.91856009|  PASSED
 marsaglia_tsang_gcd|   0|  10000000|     100|0.29619428|  PASSED
 marsaglia_tsang_gcd|   0|  10000000|     100|0.75426874|  PASSED
         sts_monobit|   1|    100000|     100|0.49959275|  PASSED
            sts_runs|   2|    100000|     100|0.78844724|  PASSED
          sts_serial|   1|    100000|     100|0.72435469|  PASSED
          sts_serial|   2|    100000|     100|0.92098160|  PASSED
          sts_serial|   3|    100000|     100|0.70195499|  PASSED
          sts_serial|   3|    100000|     100|0.57071891|  PASSED
          sts_serial|   4|    100000|     100|0.47091269|  PASSED
          sts_serial|   4|    100000|     100|0.24626278|  PASSED
          sts_serial|   5|    100000|     100|0.99669531|   WEAK
          sts_serial|   5|    100000|     100|0.93881982|  PASSED
          sts_serial|   6|    100000|     100|0.79976628|  PASSED
          sts_serial|   6|    100000|     100|0.54952100|  PASSED
          sts_serial|   7|    100000|     100|0.72445145|  PASSED
          sts_serial|   7|    100000|     100|0.99746027|   WEAK
          sts_serial|   8|    100000|     100|0.69463917|  PASSED
          sts_serial|   8|    100000|     100|0.34859315|  PASSED
          sts_serial|   9|    100000|     100|0.65658857|  PASSED
          sts_serial|   9|    100000|     100|0.82680540|  PASSED
          sts_serial|  10|    100000|     100|0.68282318|  PASSED
          sts_serial|  10|    100000|     100|0.61058368|  PASSED
          sts_serial|  11|    100000|     100|0.71861718|  PASSED
          sts_serial|  11|    100000|     100|0.89515493|  PASSED
          sts_serial|  12|    100000|     100|0.74148706|  PASSED
          sts_serial|  12|    100000|     100|0.26932113|  PASSED
          sts_serial|  13|    100000|     100|0.99949313|   WEAK
          sts_serial|  13|    100000|     100|0.93933919|  PASSED
          sts_serial|  14|    100000|     100|0.59969741|  PASSED
          sts_serial|  14|    100000|     100|0.29776462|  PASSED
          sts_serial|  15|    100000|     100|0.47338425|  PASSED
          sts_serial|  15|    100000|     100|0.16159427|  PASSED
          sts_serial|  16|    100000|     100|0.97274234|  PASSED
          sts_serial|  16|    100000|     100|0.51845614|  PASSED
         rgb_bitdist|   1|    100000|     100|0.97282348|  PASSED
         rgb_bitdist|   2|    100000|     100|0.08043995|  PASSED
         rgb_bitdist|   3|    100000|     100|0.28874219|  PASSED
         rgb_bitdist|   4|    100000|     100|0.13818700|  PASSED
         rgb_bitdist|   5|    100000|     100|0.40809359|  PASSED
         rgb_bitdist|   6|    100000|     100|0.90206433|  PASSED
         rgb_bitdist|   7|    100000|     100|0.34021584|  PASSED
```

```
        rgb_bitdist|   8|    100000|     100|0.81531523|  PASSED
        rgb_bitdist|   9|    100000|     100|0.23865448|  PASSED
        rgb_bitdist|  10|    100000|     100|0.84465684|  PASSED
        rgb_bitdist|  11|    100000|     100|0.47399477|  PASSED
        rgb_bitdist|  12|    100000|     100|0.16549648|  PASSED
rgb_minimum_distance|   2|     10000|    1000|0.12250151|  PASSED
rgb_minimum_distance|   3|     10000|    1000|0.88632161|  PASSED
rgb_minimum_distance|   4|     10000|    1000|0.84267001|  PASSED
rgb_minimum_distance|   5|     10000|    1000|0.05989242|  PASSED
    rgb_permutations|   2|    100000|     100|0.49459444|  PASSED
    rgb_permutations|   3|    100000|     100|0.36762449|  PASSED
    rgb_permutations|   4|    100000|     100|0.52682302|  PASSED
    rgb_permutations|   5|    100000|     100|0.32825974|  PASSED
      rgb_lagged_sum|   0|   1000000|     100|0.45680694|  PASSED
      rgb_lagged_sum|   1|   1000000|     100|0.54564159|  PASSED
      rgb_lagged_sum|   2|   1000000|     100|0.70765046|  PASSED
      rgb_lagged_sum|   3|   1000000|     100|0.79892901|  PASSED
      rgb_lagged_sum|   4|   1000000|     100|0.62202415|  PASSED
      rgb_lagged_sum|   5|   1000000|     100|0.82119915|  PASSED
      rgb_lagged_sum|   6|   1000000|     100|0.89224820|  PASSED
      rgb_lagged_sum|   7|   1000000|     100|0.54912437|  PASSED
      rgb_lagged_sum|   8|   1000000|     100|0.20879414|  PASSED
      rgb_lagged_sum|   9|   1000000|     100|0.06459332|  PASSED
      rgb_lagged_sum|  10|   1000000|     100|0.54787327|  PASSED
      rgb_lagged_sum|  11|   1000000|     100|0.87919812|  PASSED
      rgb_lagged_sum|  12|   1000000|     100|0.75681773|  PASSED
      rgb_lagged_sum|  13|   1000000|     100|0.11763166|  PASSED
      rgb_lagged_sum|  14|   1000000|     100|0.73742242|  PASSED
      rgb_lagged_sum|  15|   1000000|     100|0.30914542|  PASSED
      rgb_lagged_sum|  16|   1000000|     100|0.58505819|  PASSED
      rgb_lagged_sum|  17|   1000000|     100|0.81390089|  PASSED
      rgb_lagged_sum|  18|   1000000|     100|0.77810421|  PASSED
      rgb_lagged_sum|  19|   1000000|     100|0.20879803|  PASSED
      rgb_lagged_sum|  20|   1000000|     100|0.47675291|  PASSED
      rgb_lagged_sum|  21|   1000000|     100|0.58019509|  PASSED
      rgb_lagged_sum|  22|   1000000|     100|0.51400827|  PASSED
      rgb_lagged_sum|  23|   1000000|     100|0.11021201|  PASSED
      rgb_lagged_sum|  24|   1000000|     100|0.40830609|  PASSED
      rgb_lagged_sum|  25|   1000000|     100|0.06650902|  PASSED
      rgb_lagged_sum|  26|   1000000|     100|0.00543357|  PASSED
      rgb_lagged_sum|  27|   1000000|     100|0.50118182|  PASSED
      rgb_lagged_sum|  28|   1000000|     100|0.86761086|  PASSED
      rgb_lagged_sum|  29|   1000000|     100|0.62221948|  PASSED
      rgb_lagged_sum|  30|   1000000|     100|0.43794640|  PASSED
      rgb_lagged_sum|  31|   1000000|     100|0.87900118|  PASSED
      rgb_lagged_sum|  32|   1000000|     100|0.26657919|  PASSED
      rgb_kstest_test|  0|     10000|    1000|0.65301741|  PASSED
       dab_bytedistrib|  0|  51200000|       1|0.60349920|  PASSED
              dab_dct| 256|    50000|       1|0.65273698|  PASSED
Preparing to run test 207.  ntuple = 0
         dab_filltree|  32|  15000000|       1|0.93852215|  PASSED
         dab_filltree|  32|  15000000|       1|0.84733656|  PASSED
Preparing to run test 208.  ntuple = 0
        dab_filltree2|   0|   5000000|       1|0.18347079|  PASSED
        dab_filltree2|   1|   5000000|       1|0.18845977|  PASSED
Preparing to run test 209.  ntuple = 0
         dab_monobit2|  12|  65000000|       1|0.87729688|  PASSED
```

# 12. APPENDIX – DIEHARDER TEST: /DEV/RANDOM

```
#=============================================================================#
#            dieharder version 3.31.1 Copyright 2003 Robert G. Brown           #
#=============================================================================#
   rng_name    |           filename             |rands/second|
 file_input_raw|                 dev-random.raw |  1.55e+07  |
#=============================================================================#
        test_name   |ntup| tsamples |psamples|  p-value |Assessment
#=============================================================================#
   diehard_birthdays|   0|       100|     100|0.56076347|  PASSED
      diehard_operm5|   0|   1000000|     100|0.54533269|  PASSED
  diehard_rank_32x32|   0|     40000|     100|0.32000993|  PASSED
    diehard_rank_6x8|   0|    100000|     100|0.76275024|  PASSED
   diehard_bitstream|   0|   2097152|     100|0.63053368|  PASSED
        diehard_opso|   0|   2097152|     100|0.12079836|  PASSED
        diehard_oqso|   0|   2097152|     100|0.45499297|  PASSED
         diehard_dna|   0|   2097152|     100|0.87265280|  PASSED
diehard_count_1s_str|   0|    256000|     100|0.28132278|  PASSED
diehard_count_1s_byt|   0|    256000|     100|0.38232403|  PASSED
 diehard_parking_lot|   0|     12000|     100|0.01002226|  PASSED
    diehard_2dsphere|   2|      8000|     100|0.92254447|  PASSED
    diehard_3dsphere|   3|      4000|     100|0.07066726|  PASSED
     diehard_squeeze|   0|    100000|     100|0.57747960|  PASSED
        diehard_sums|   0|       100|     100|0.93029416|  PASSED
        diehard_runs|   0|    100000|     100|0.18406317|  PASSED
        diehard_runs|   0|    100000|     100|0.54953016|  PASSED
       diehard_craps|   0|    200000|     100|0.86703027|  PASSED
       diehard_craps|   0|    200000|     100|0.30639650|  PASSED
 marsaglia_tsang_gcd|   0|  10000000|     100|0.14538809|  PASSED
 marsaglia_tsang_gcd|   0|  10000000|     100|0.95971817|  PASSED
         sts_monobit|   1|    100000|     100|0.81392856|  PASSED
            sts_runs|   2|    100000|     100|0.99937546|   WEAK
          sts_serial|   1|    100000|     100|0.82060740|  PASSED
          sts_serial|   2|    100000|     100|0.67456120|  PASSED
          sts_serial|   3|    100000|     100|0.92967246|  PASSED
          sts_serial|   3|    100000|     100|0.94029458|  PASSED
          sts_serial|   4|    100000|     100|0.98618610|  PASSED
          sts_serial|   4|    100000|     100|0.99404514|  PASSED
          sts_serial|   5|    100000|     100|0.50281719|  PASSED
          sts_serial|   5|    100000|     100|0.43512984|  PASSED
          sts_serial|   6|    100000|     100|0.66359687|  PASSED
          sts_serial|   6|    100000|     100|0.33434014|  PASSED
          sts_serial|   7|    100000|     100|0.36787347|  PASSED
          sts_serial|   7|    100000|     100|0.82204473|  PASSED
          sts_serial|   8|    100000|     100|0.88923566|  PASSED
          sts_serial|   8|    100000|     100|0.45407924|  PASSED
          sts_serial|   9|    100000|     100|0.21567827|  PASSED
          sts_serial|   9|    100000|     100|0.37497504|  PASSED
          sts_serial|  10|    100000|     100|0.76400076|  PASSED
          sts_serial|  10|    100000|     100|0.53647825|  PASSED
          sts_serial|  11|    100000|     100|0.97956255|  PASSED
          sts_serial|  11|    100000|     100|0.37232324|  PASSED
          sts_serial|  12|    100000|     100|0.27398505|  PASSED
          sts_serial|  12|    100000|     100|0.38751517|  PASSED
          sts_serial|  13|    100000|     100|0.26127727|  PASSED
          sts_serial|  13|    100000|     100|0.30868235|  PASSED
          sts_serial|  14|    100000|     100|0.60608777|  PASSED
          sts_serial|  14|    100000|     100|0.83148107|  PASSED
          sts_serial|  15|    100000|     100|0.67152129|  PASSED
          sts_serial|  15|    100000|     100|0.71528346|  PASSED
          sts_serial|  16|    100000|     100|0.81202823|  PASSED
          sts_serial|  16|    100000|     100|0.52628960|  PASSED
         rgb_bitdist|   1|    100000|     100|0.60189355|  PASSED
         rgb_bitdist|   2|    100000|     100|0.99769636|   WEAK
         rgb_bitdist|   3|    100000|     100|0.93347342|  PASSED
         rgb_bitdist|   4|    100000|     100|0.89292777|  PASSED
         rgb_bitdist|   5|    100000|     100|0.05387356|  PASSED
         rgb_bitdist|   6|    100000|     100|0.20423878|  PASSED
         rgb_bitdist|   7|    100000|     100|0.97060945|  PASSED
         rgb_bitdist|   8|    100000|     100|0.42237803|  PASSED
         rgb_bitdist|   9|    100000|     100|0.06869971|  PASSED
```

```
        rgb_bitdist|  10|    100000|     100|0.90974564|  PASSED
        rgb_bitdist|  11|    100000|     100|0.50820739|  PASSED
        rgb_bitdist|  12|    100000|     100|0.82809420|  PASSED
rgb_minimum_distance|   2|     10000|    1000|0.28472200|  PASSED
rgb_minimum_distance|   3|     10000|    1000|0.76366305|  PASSED
rgb_minimum_distance|   4|     10000|    1000|0.96187766|  PASSED
rgb_minimum_distance|   5|     10000|    1000|0.08315055|  PASSED
    rgb_permutations|   2|    100000|     100|0.85229696|  PASSED
    rgb_permutations|   3|    100000|     100|0.77998375|  PASSED
    rgb_permutations|   4|    100000|     100|0.86253038|  PASSED
    rgb_permutations|   5|    100000|     100|0.39147614|  PASSED
      rgb_lagged_sum|   0|   1000000|     100|0.67004506|  PASSED
      rgb_lagged_sum|   1|   1000000|     100|0.86909895|  PASSED
      rgb_lagged_sum|   2|   1000000|     100|0.50354994|  PASSED
      rgb_lagged_sum|   3|   1000000|     100|0.18308552|  PASSED
      rgb_lagged_sum|   4|   1000000|     100|0.74084618|  PASSED
      rgb_lagged_sum|   5|   1000000|     100|0.69356258|  PASSED
      rgb_lagged_sum|   6|   1000000|     100|0.04038407|  PASSED
      rgb_lagged_sum|   7|   1000000|     100|0.14955176|  PASSED
      rgb_lagged_sum|   8|   1000000|     100|0.33254496|  PASSED
      rgb_lagged_sum|   9|   1000000|     100|0.11220245|  PASSED
      rgb_lagged_sum|  10|   1000000|     100|0.02550530|  PASSED
      rgb_lagged_sum|  11|   1000000|     100|0.94292634|  PASSED
      rgb_lagged_sum|  12|   1000000|     100|0.75302333|  PASSED
      rgb_lagged_sum|  13|   1000000|     100|0.18376902|  PASSED
      rgb_lagged_sum|  14|   1000000|     100|0.67671462|  PASSED
      rgb_lagged_sum|  15|   1000000|     100|0.78142945|  PASSED
      rgb_lagged_sum|  16|   1000000|     100|0.29636210|  PASSED
      rgb_lagged_sum|  17|   1000000|     100|0.55623898|  PASSED
      rgb_lagged_sum|  18|   1000000|     100|0.97525959|  PASSED
      rgb_lagged_sum|  19|   1000000|     100|0.41378251|  PASSED
      rgb_lagged_sum|  20|   1000000|     100|0.14589978|  PASSED
      rgb_lagged_sum|  21|   1000000|     100|0.62827593|  PASSED
      rgb_lagged_sum|  22|   1000000|     100|0.71224169|  PASSED
      rgb_lagged_sum|  23|   1000000|     100|0.63029963|  PASSED
      rgb_lagged_sum|  24|   1000000|     100|0.75292718|  PASSED
      rgb_lagged_sum|  25|   1000000|     100|0.79309876|  PASSED
      rgb_lagged_sum|  26|   1000000|     100|0.22623464|  PASSED
      rgb_lagged_sum|  27|   1000000|     100|0.25957095|  PASSED
      rgb_lagged_sum|  28|   1000000|     100|0.18866673|  PASSED
      rgb_lagged_sum|  29|   1000000|     100|0.02690510|  PASSED
      rgb_lagged_sum|  30|   1000000|     100|0.56285030|  PASSED
      rgb_lagged_sum|  31|   1000000|     100|0.44184137|  PASSED
      rgb_lagged_sum|  32|   1000000|     100|0.99892846|    WEAK
      rgb_kstest_test|  0|     10000|    1000|0.97243855|  PASSED
       dab_bytedistrib|  0|  51200000|       1|0.10844598|  PASSED
              dab_dct| 256|    50000|       1|0.37095973|  PASSED
Preparing to run test 207.  ntuple = 0
         dab_filltree|  32|  15000000|       1|0.28905522|  PASSED
         dab_filltree|  32|  15000000|       1|0.36484758|  PASSED
Preparing to run test 208.  ntuple = 0
        dab_filltree2|   0|   5000000|       1|0.51053208|  PASSED
        dab_filltree2|   1|   5000000|       1|0.97084031|  PASSED
Preparing to run test 209.  ntuple = 0
         dab_monobit2|  12|  65000000|       1|0.94701176|  PASSED
```

# 13. APPENDIX – DIEHARDER TEST: PI BUILT-IN

```
$ more dieharder-test-pi-builtin-random.txt
#=============================================================================#
#            dieharder version 3.31.1 Copyright 2003 Robert G. Brown          #
#=============================================================================#
   rng_name    |           filename             |rands/second|
 file_input_raw|        pi-builtin-random.raw|  1.05e+07  |
#=============================================================================#
        test_name   |ntup| tsamples |psamples|  p-value |Assessment
#=============================================================================#
   diehard_birthdays|   0|       100|     100|0.28688816|  PASSED
      diehard_operm5|   0|   1000000|     100|0.71215185|  PASSED
  diehard_rank_32x32|   0|     40000|     100|0.26006665|  PASSED
    diehard_rank_6x8|   0|    100000|     100|0.41525526|  PASSED
   diehard_bitstream|   0|   2097152|     100|0.40545306|  PASSED
        diehard_opso|   0|   2097152|     100|0.89107541|  PASSED
        diehard_oqso|   0|   2097152|     100|0.73216666|  PASSED
         diehard_dna|   0|   2097152|     100|0.48658022|  PASSED
diehard_count_1s_str|   0|    256000|     100|0.13335067|  PASSED
diehard_count_1s_byt|   0|    256000|     100|0.74042755|  PASSED
 diehard_parking_lot|   0|     12000|     100|0.59033501|  PASSED
    diehard_2dsphere|   2|      8000|     100|0.97594943|  PASSED
    diehard_3dsphere|   3|      4000|     100|0.62708217|  PASSED
     diehard_squeeze|   0|    100000|     100|0.38140481|  PASSED
        diehard_sums|   0|       100|     100|0.03294898|  PASSED
        diehard_runs|   0|    100000|     100|0.01743073|  PASSED
        diehard_runs|   0|    100000|     100|0.74448223|  PASSED
       diehard_craps|   0|    200000|     100|0.87703388|  PASSED
       diehard_craps|   0|    200000|     100|0.84304348|  PASSED
 marsaglia_tsang_gcd|   0|  10000000|     100|0.19678462|  PASSED
 marsaglia_tsang_gcd|   0|  10000000|     100|0.60607225|  PASSED
         sts_monobit|   1|    100000|     100|0.99933794|   WEAK
            sts_runs|   2|    100000|     100|0.03930543|  PASSED
          sts_serial|   1|    100000|     100|0.80693911|  PASSED
          sts_serial|   2|    100000|     100|0.39789799|  PASSED
          sts_serial|   3|    100000|     100|0.90101190|  PASSED
          sts_serial|   3|    100000|     100|0.68484986|  PASSED
          sts_serial|   4|    100000|     100|0.23118460|  PASSED
          sts_serial|   4|    100000|     100|0.66038884|  PASSED
          sts_serial|   5|    100000|     100|0.52591780|  PASSED
          sts_serial|   5|    100000|     100|0.77791272|  PASSED
          sts_serial|   6|    100000|     100|0.63472361|  PASSED
          sts_serial|   6|    100000|     100|0.87919339|  PASSED
          sts_serial|   7|    100000|     100|0.76668178|  PASSED
          sts_serial|   7|    100000|     100|0.45321114|  PASSED
          sts_serial|   8|    100000|     100|0.22874552|  PASSED
          sts_serial|   8|    100000|     100|0.24832635|  PASSED
          sts_serial|   9|    100000|     100|0.01406337|  PASSED
          sts_serial|   9|    100000|     100|0.01477669|  PASSED
          sts_serial|  10|    100000|     100|0.07691219|  PASSED
          sts_serial|  10|    100000|     100|0.68904429|  PASSED
          sts_serial|  11|    100000|     100|0.97067835|  PASSED
          sts_serial|  11|    100000|     100|0.19196503|  PASSED
          sts_serial|  12|    100000|     100|0.89192782|  PASSED
          sts_serial|  12|    100000|     100|0.45426198|  PASSED
          sts_serial|  13|    100000|     100|0.61261279|  PASSED
          sts_serial|  13|    100000|     100|0.92786301|  PASSED
          sts_serial|  14|    100000|     100|0.81208829|  PASSED
          sts_serial|  14|    100000|     100|0.75406286|  PASSED
          sts_serial|  15|    100000|     100|0.79132046|  PASSED
          sts_serial|  15|    100000|     100|0.70659001|  PASSED
          sts_serial|  16|    100000|     100|0.92964638|  PASSED
          sts_serial|  16|    100000|     100|0.93632297|  PASSED
         rgb_bitdist|   1|    100000|     100|0.96987845|  PASSED
         rgb_bitdist|   2|    100000|     100|0.97177581|  PASSED
         rgb_bitdist|   3|    100000|     100|0.97010290|  PASSED
         rgb_bitdist|   4|    100000|     100|0.27709112|  PASSED
         rgb_bitdist|   5|    100000|     100|0.64107279|  PASSED
         rgb_bitdist|   6|    100000|     100|0.35094460|  PASSED
         rgb_bitdist|   7|    100000|     100|0.31953730|  PASSED
         rgb_bitdist|   8|    100000|     100|0.37452958|  PASSED
```

```
        rgb_bitdist|   9|    100000|     100|0.18606734|  PASSED
        rgb_bitdist|  10|    100000|     100|0.44688654|  PASSED
        rgb_bitdist|  11|    100000|     100|0.98930489|  PASSED
        rgb_bitdist|  12|    100000|     100|0.71400461|  PASSED
rgb_minimum_distance|   2|     10000|    1000|0.55038886|  PASSED
rgb_minimum_distance|   3|     10000|    1000|0.58175990|  PASSED
rgb_minimum_distance|   4|     10000|    1000|0.19573296|  PASSED
rgb_minimum_distance|   5|     10000|    1000|0.82291677|  PASSED
    rgb_permutations|   2|    100000|     100|0.97564438|  PASSED
    rgb_permutations|   3|    100000|     100|0.95561474|  PASSED
    rgb_permutations|   4|    100000|     100|0.46812640|  PASSED
    rgb_permutations|   5|    100000|     100|0.12999627|  PASSED
      rgb_lagged_sum|   0|   1000000|     100|0.60526096|  PASSED
      rgb_lagged_sum|   1|   1000000|     100|0.74907041|  PASSED
      rgb_lagged_sum|   2|   1000000|     100|0.14046130|  PASSED
      rgb_lagged_sum|   3|   1000000|     100|0.84367599|  PASSED
      rgb_lagged_sum|   4|   1000000|     100|0.55208422|  PASSED
      rgb_lagged_sum|   5|   1000000|     100|0.92752218|  PASSED
      rgb_lagged_sum|   6|   1000000|     100|0.53848595|  PASSED
      rgb_lagged_sum|   7|   1000000|     100|0.52446225|  PASSED
      rgb_lagged_sum|   8|   1000000|     100|0.07821347|  PASSED
      rgb_lagged_sum|   9|   1000000|     100|0.72865610|  PASSED
      rgb_lagged_sum|  10|   1000000|     100|0.79467779|  PASSED
      rgb_lagged_sum|  11|   1000000|     100|0.96264475|  PASSED
      rgb_lagged_sum|  12|   1000000|     100|0.87741678|  PASSED
      rgb_lagged_sum|  13|   1000000|     100|0.69901222|  PASSED
      rgb_lagged_sum|  14|   1000000|     100|0.39866587|  PASSED
      rgb_lagged_sum|  15|   1000000|     100|0.97582528|  PASSED
      rgb_lagged_sum|  16|   1000000|     100|0.64930785|  PASSED
      rgb_lagged_sum|  17|   1000000|     100|0.36487658|  PASSED
      rgb_lagged_sum|  18|   1000000|     100|0.32561193|  PASSED
      rgb_lagged_sum|  19|   1000000|     100|0.56034261|  PASSED
      rgb_lagged_sum|  20|   1000000|     100|0.54579125|  PASSED
      rgb_lagged_sum|  21|   1000000|     100|0.13104318|  PASSED
      rgb_lagged_sum|  22|   1000000|     100|0.97371703|  PASSED
      rgb_lagged_sum|  23|   1000000|     100|0.22569577|  PASSED
      rgb_lagged_sum|  24|   1000000|     100|0.99885183|    WEAK
      rgb_lagged_sum|  25|   1000000|     100|0.49187439|  PASSED
      rgb_lagged_sum|  26|   1000000|     100|0.91917567|  PASSED
      rgb_lagged_sum|  27|   1000000|     100|0.15712840|  PASSED
      rgb_lagged_sum|  28|   1000000|     100|0.10931251|  PASSED
      rgb_lagged_sum|  29|   1000000|     100|0.59007041|  PASSED
      rgb_lagged_sum|  30|   1000000|     100|0.56567619|  PASSED
      rgb_lagged_sum|  31|   1000000|     100|0.45601290|  PASSED
      rgb_lagged_sum|  32|   1000000|     100|0.54987187|  PASSED
      rgb_kstest_test|  0|     10000|    1000|0.20895553|  PASSED
       dab_bytedistrib|  0|  51200000|       1|0.40280744|  PASSED
               dab_dct|256|     50000|       1|0.88542539|  PASSED
Preparing to run test 207.  ntuple = 0
          dab_filltree| 32|  15000000|       1|0.65730333|  PASSED
          dab_filltree| 32|  15000000|       1|0.59811659|  PASSED
Preparing to run test 208.  ntuple = 0
         dab_filltree2|  0|   5000000|       1|0.57733908|  PASSED
         dab_filltree2|  1|   5000000|       1|0.46380463|  PASSED
Preparing to run test 209.  ntuple = 0
          dab_monobit2| 12|  65000000|       1|0.65444480|  PASSED
```

# 14. APPENDIX – DIEHARDER TEST: NEUG

```
$ more dieharder-test-neug-random.txt
#=============================================================================#
#            dieharder version 3.31.1 Copyright 2003 Robert G. Brown          #
#=============================================================================#
   rng_name    |           filename             |rands/second|
 file_input_raw|                 neug-random.raw|  1.61e+07  |
#=============================================================================#
        test_name   |ntup| tsamples |psamples|  p-value |Assessment
#=============================================================================#
   diehard_birthdays|   0|       100|     100|0.02720373|  PASSED
      diehard_operm5|   0|   1000000|     100|0.50594413|  PASSED
  diehard_rank_32x32|   0|     40000|     100|0.80044834|  PASSED
    diehard_rank_6x8|   0|    100000|     100|0.56116610|  PASSED
   diehard_bitstream|   0|   2097152|     100|0.83370023|  PASSED
        diehard_opso|   0|   2097152|     100|0.95564329|  PASSED
        diehard_oqso|   0|   2097152|     100|0.46410706|  PASSED
         diehard_dna|   0|   2097152|     100|0.73965110|  PASSED
diehard_count_1s_str|   0|    256000|     100|0.18643691|  PASSED
diehard_count_1s_byt|   0|    256000|     100|0.26398686|  PASSED
 diehard_parking_lot|   0|     12000|     100|0.54444845|  PASSED
    diehard_2dsphere|   2|      8000|     100|0.78962307|  PASSED
    diehard_3dsphere|   3|      4000|     100|0.90519655|  PASSED
     diehard_squeeze|   0|    100000|     100|0.86902967|  PASSED
        diehard_sums|   0|       100|     100|0.32788953|  PASSED
        diehard_runs|   0|    100000|     100|0.01293063|  PASSED
        diehard_runs|   0|    100000|     100|0.97162844|  PASSED
       diehard_craps|   0|    200000|     100|0.91431394|  PASSED
       diehard_craps|   0|    200000|     100|0.77463832|  PASSED
  marsaglia_tsang_gcd|  0|  10000000|     100|0.71741763|  PASSED
  marsaglia_tsang_gcd|  0|  10000000|     100|0.19844569|  PASSED
         sts_monobit|   1|    100000|     100|0.89076205|  PASSED
            sts_runs|   2|    100000|     100|0.03970224|  PASSED
          sts_serial|   1|    100000|     100|0.85303566|  PASSED
          sts_serial|   2|    100000|     100|0.98827844|  PASSED
          sts_serial|   3|    100000|     100|0.63604337|  PASSED
          sts_serial|   3|    100000|     100|0.27333066|  PASSED
          sts_serial|   4|    100000|     100|0.14008852|  PASSED
          sts_serial|   4|    100000|     100|0.21049589|  PASSED
          sts_serial|   5|    100000|     100|0.17750107|  PASSED
          sts_serial|   5|    100000|     100|0.35868297|  PASSED
          sts_serial|   6|    100000|     100|0.31621427|  PASSED
          sts_serial|   6|    100000|     100|0.73220832|  PASSED
          sts_serial|   7|    100000|     100|0.47255193|  PASSED
          sts_serial|   7|    100000|     100|0.47464138|  PASSED
          sts_serial|   8|    100000|     100|0.14069505|  PASSED
          sts_serial|   8|    100000|     100|0.66506616|  PASSED
          sts_serial|   9|    100000|     100|0.91487528|  PASSED
          sts_serial|   9|    100000|     100|0.88166035|  PASSED
          sts_serial|  10|    100000|     100|0.09767396|  PASSED
          sts_serial|  10|    100000|     100|0.15594996|  PASSED
          sts_serial|  11|    100000|     100|0.32901497|  PASSED
          sts_serial|  11|    100000|     100|0.82795317|  PASSED
          sts_serial|  12|    100000|     100|0.39266048|  PASSED
          sts_serial|  12|    100000|     100|0.72305874|  PASSED
          sts_serial|  13|    100000|     100|0.46249765|  PASSED
          sts_serial|  13|    100000|     100|0.96949949|  PASSED
          sts_serial|  14|    100000|     100|0.99004878|  PASSED
          sts_serial|  14|    100000|     100|0.08325366|  PASSED
          sts_serial|  15|    100000|     100|0.68517592|  PASSED
          sts_serial|  15|    100000|     100|0.81872374|  PASSED
          sts_serial|  16|    100000|     100|0.93019224|  PASSED
          sts_serial|  16|    100000|     100|0.88184439|  PASSED
         rgb_bitdist|   1|    100000|     100|0.86207305|  PASSED
         rgb_bitdist|   2|    100000|     100|0.19177094|  PASSED
         rgb_bitdist|   3|    100000|     100|0.95847307|  PASSED
         rgb_bitdist|   4|    100000|     100|0.53843358|  PASSED
         rgb_bitdist|   5|    100000|     100|0.76616687|  PASSED
         rgb_bitdist|   6|    100000|     100|0.83609469|  PASSED
         rgb_bitdist|   7|    100000|     100|0.73276465|  PASSED
         rgb_bitdist|   8|    100000|     100|0.06172539|  PASSED
```

```
      rgb_bitdist|   9|    100000|     100|0.46369493|  PASSED
      rgb_bitdist|  10|    100000|     100|0.83278715|  PASSED
      rgb_bitdist|  11|    100000|     100|0.58550503|  PASSED
      rgb_bitdist|  12|    100000|     100|0.42537279|  PASSED
rgb_minimum_distance|   2|     10000|    1000|0.63673038|  PASSED
rgb_minimum_distance|   3|     10000|    1000|0.79580054|  PASSED
rgb_minimum_distance|   4|     10000|    1000|0.99482281|  PASSED
rgb_minimum_distance|   5|     10000|    1000|0.73252922|  PASSED
   rgb_permutations|   2|    100000|     100|0.93102570|  PASSED
   rgb_permutations|   3|    100000|     100|0.99196479|  PASSED
   rgb_permutations|   4|    100000|     100|0.99794238|    WEAK
   rgb_permutations|   5|    100000|     100|0.70583102|  PASSED
     rgb_lagged_sum|   0|   1000000|     100|0.89911516|  PASSED
     rgb_lagged_sum|   1|   1000000|     100|0.81100514|  PASSED
     rgb_lagged_sum|   2|   1000000|     100|0.71454619|  PASSED
     rgb_lagged_sum|   3|   1000000|     100|0.76150327|  PASSED
     rgb_lagged_sum|   4|   1000000|     100|0.78755611|  PASSED
     rgb_lagged_sum|   5|   1000000|     100|0.12453920|  PASSED
     rgb_lagged_sum|   6|   1000000|     100|0.72041832|  PASSED
     rgb_lagged_sum|   7|   1000000|     100|0.75752579|  PASSED
     rgb_lagged_sum|   8|   1000000|     100|0.31352444|  PASSED
     rgb_lagged_sum|   9|   1000000|     100|0.73537387|  PASSED
     rgb_lagged_sum|  10|   1000000|     100|0.87852817|  PASSED
     rgb_lagged_sum|  11|   1000000|     100|0.93726342|  PASSED
     rgb_lagged_sum|  12|   1000000|     100|0.85749941|  PASSED
     rgb_lagged_sum|  13|   1000000|     100|0.80435141|  PASSED
     rgb_lagged_sum|  14|   1000000|     100|0.54617316|  PASSED
     rgb_lagged_sum|  15|   1000000|     100|0.27230666|  PASSED
     rgb_lagged_sum|  16|   1000000|     100|0.59010132|  PASSED
     rgb_lagged_sum|  17|   1000000|     100|0.44808597|  PASSED
     rgb_lagged_sum|  18|   1000000|     100|0.26119680|  PASSED
     rgb_lagged_sum|  19|   1000000|     100|0.38892342|  PASSED
     rgb_lagged_sum|  20|   1000000|     100|0.66053096|  PASSED
     rgb_lagged_sum|  21|   1000000|     100|0.86246821|  PASSED
     rgb_lagged_sum|  22|   1000000|     100|0.98146311|  PASSED
     rgb_lagged_sum|  23|   1000000|     100|0.21185090|  PASSED
     rgb_lagged_sum|  24|   1000000|     100|0.34542950|  PASSED
     rgb_lagged_sum|  25|   1000000|     100|0.81893495|  PASSED
     rgb_lagged_sum|  26|   1000000|     100|0.88053540|  PASSED
     rgb_lagged_sum|  27|   1000000|     100|0.03652345|  PASSED
     rgb_lagged_sum|  28|   1000000|     100|0.10270837|  PASSED
     rgb_lagged_sum|  29|   1000000|     100|0.93413152|  PASSED
     rgb_lagged_sum|  30|   1000000|     100|0.10521408|  PASSED
     rgb_lagged_sum|  31|   1000000|     100|0.74603708|  PASSED
     rgb_lagged_sum|  32|   1000000|     100|0.99856776|    WEAK
     rgb_kstest_test|   0|     10000|    1000|0.96149249|  PASSED
     dab_bytedistrib|   0|  51200000|       1|0.45216823|  PASSED
             dab_dct| 256|     50000|       1|0.49959599|  PASSED
Preparing to run test 207.  ntuple = 0
        dab_filltree|  32|  15000000|       1|0.03935264|  PASSED
        dab_filltree|  32|  15000000|       1|0.94047345|  PASSED
Preparing to run test 208.  ntuple = 0
       dab_filltree2|   0|   5000000|       1|0.69101210|  PASSED
       dab_filltree2|   1|   5000000|       1|0.33741247|  PASSED
Preparing to run test 209.  ntuple = 0
        dab_monobit2|  12|  65000000|       1|0.41345852|  PASSED
```